\input harvmac
\input epsf.sty


\noblackbox

\font\cmss=cmss10
\font\cmsss=cmss10 at 7pt

\def\inbar{\vrule height1.5ex width.4pt depth0pt}

\def\IN{\relax{\rm I\kern-.18em N}}
\def\IB{\relax\hbox{$\inbar\kern-.3em{\rm B}$}}
\def\IC{\relax\hbox{$\inbar\kern-.3em{\rm C}$}}
\def\IQ{\relax\hbox{$\inbar\kern-.3em{\rm Q}$}}
\def\ID{\relax\hbox{$\inbar\kern-.3em{\rm D}$}}
\def\IE{\relax\hbox{$\inbar\kern-.3em{\rm E}$}}
\def\IF{\relax\hbox{$\inbar\kern-.3em{\rm F}$}}
\def\IG{\relax\hbox{$\inbar\kern-.3em{\rm G}$}}
\def\IGa{\relax\hbox{${\rm I}\kern-.18em\Gamma$}}
\def\IH{\relax{\rm I\kern-.18em H}}
\def\IK{\relax{\rm I\kern-.18em K}}
\def\IL{\relax{\rm I\kern-.18em L}}
\def\IP{\relax{\rm I\kern-.18em P}}
\def\IR{\relax{\rm I\kern-.18em R}}
\def\Z{\relax\ifmmode\mathchoice{\hbox{\cmss Z\kern-.4em Z}}{\hbox{\cmss Z\kern-.4em Z}} {\lower.9pt\hbox{\cmsss Z\kern-.4em Z}}{\lower1.2pt\hbox{\cmsss Z\kern-.4em Z}}\else{\cmss Z\kern-.4em Z}\fi}

\def\II{\relax{\rm I\kern-.18em I}}
\def\one{\relax{\rm 1\kern-.25em I}}

\def\S{Sec.~}

\def\CLL{\relax{\CL\kern-.74em \CL}}

\def\sech{{\rm sech}}

\def\sgn{{\rm sgn}}


\lref\fermiondoubling{ 
H. B. Nielsen and M. Ninomiya,
``Absence of Neutrinos on a Lattice. 1. Proof by Homotopy Theory,"
 Nucl.\ Phys.\  B {\bf 185}, 20 (1981).
}

\lref\SternWeak{
Zohar Ringel  and Yaacov E. Kraus and Ady Stern,
``Strong side of weak topological insulators",
Phys.\ Rev.\  B {\bf 86}, 045102 (2012).
}

\lref\fukanetopins{
L.~Fu and C.~L.~ Kane,
``Topological insulators with inversion symmetry,"
Phys. Rev. B {\bf 76}, 045302 (2007).
}

\lref\FuKanePump{
L.~ Fu and C.~L. Kane,
``Time reversal polarization and a Z2 adiabatic spin pump,"
Phys.\ Rev.\  B {\bf 74}, 195312 (2006).
}

\lref\RyuMooreLudwig{
S. Ryu and J. Moore and A. W. W. Ludwig,
``Electromagnetic and gravitational responses and anomalies in topological insulators and superconductors,"
Phys.\ Rev.\  B {\bf 85}, 045104 (2012).
}

\lref\MaciejkoTX{
  J.~Maciejko, X.~-L.~Qi, A.~Karch and S.~-C.~Zhang,
  ``Fractional topological insulators in three dimensions,''
Phys.\ Rev.\ Lett.\  {\bf 105}, 246809 (2010).
[arXiv:1004.3628 [cond-mat.str-el]].
}

\lref\SwingleEtAl{
B. Swingle and M Barkeshli and J. McGreevy and T. Senthil, 
``Correlated topological insulators and the fractional magnetoelectric effect,"
Phys.\ Rev.\  B {\bf 83}, 0195139 (2011).
}

\lref\MacjeckoLong{
Joseph Maciejko and Xiao-Liang Qi and Andreas Karch and Shou-Cheng Zhang, 
``Models of three-dimensional fractional topological insulators",
arXiv:1111.6816. 
}

\lref\LevBur{
 Michael Levin and F. J. Burnell and Maciej  Koch-Janusz and Ady Stern, 
``Exactly soluble models for fractional topological insulators in two and three dimensions",
 Phys.\ Rev.\  B {\bf 84}, 235145 (2011).
}

\lref\shankar{
R. Shankar,
``Renormalization-group approach to interacting fermions,"
Rev. Mod. Phys. {\bf 66}, 129 (1994).
}

\lref\JackiwFN{
  R.~Jackiw and C.~Rebbi,
  ``Solitons With Fermion Number 1/2,"
  Phys.\ Rev.\  D {\bf 13}, 3398 (1976).
}

\lref\RedlichKN{
  A.~N.~Redlich,
  ``Gauge Non-Invariance and Parity Non-Conservation of Three-Dimensional Fermions,"
  Phys.\ Rev.\ Lett.\  {\bf 52}, 18 (1984).
}

\lref\RedlichDV{
  A.~N.~Redlich,
  ``Parity Violation And Gauge Noninvariance Of The Effective Gauge Field
  Action In Three-Dimensions,''
  Phys.\ Rev.\  D {\bf 29}, 2366 (1984).
}

\lref\AlvarezGaumeIG{
  L.~Alvarez-Gaume and E.~Witten,
  ``Gravitational Anomalies,''
  Nucl.\ Phys.\  B {\bf 234}, 269 (1984).
}

\lref\CallanSA{
  C.~G.~Callan and J.~A.~Harvey,
  ``Anomalies And Fermion Zero Modes On Strings And Domain Walls,''
  Nucl.\ Phys.\  B {\bf 250}, 427 (1985).
}

\lref\GoldstoneKK{
  J.~Goldstone and F.~Wilczek,
  ``Fractional Quantum Numbers On Solitons,''
  Phys.\ Rev.\ Lett.\  {\bf 47}, 986 (1981).
}

\lref\StoneIW{
  M.~Stone,
  ``Edge Waves In The Quantum Hall Effect,''
Annals Phys.\  {\bf 207}, 38 (1991).
}

\lref\qhz{
Xiao-Liang Qi, Taylor L. Hughes, and Shou-Cheng Zhang,
``Topological field theory of time-reversal invariant insulators,"
Phys. Rev. B {\bf 78}, 195424 (2008).
}

\lref\fukane{
Liang Fu, C.L. Kane,
``Topological insulators with inversion symmetry,"
Phys. Rev. B {\bf 76}, 045302 (2007).
}

\lref\dhlee{
Dung-Hai Lee,
``Surface States of Topological Insulators: The Dirac Fermion in Curved Two-Dimensional Spaces,"
Phys. Rev. Lett. {\bf 103}, 196804 (2009).
}

\lref\AppelquistVG{
  T.~Appelquist and R.~D.~Pisarski,
  ``High-Temperature Yang-Mills Theories And Three-Dimensional Quantum
  Chromodynamics,''
  Phys.\ Rev.\  D {\bf 23}, 2305 (1981).
}

\lref\JackiwKV{
  R.~Jackiw and S.~Templeton,
  ``How Superrenormalizable Interactions Cure Their Infrared Divergences,''
  Phys.\ Rev.\  D {\bf 23}, 2291 (1981).
}

\lref\AppelquistSR{
  T.~Appelquist, D.~Nash and L.~C.~R.~Wijewardhana,
  ``Critical Behavior In (2+1)-Dimensional QED,''
  Phys.\ Rev.\ Lett.\  {\bf 60}, 2575 (1988).
}

\lref\AppelquistSF{
  T.~Appelquist and U.~W.~Heinz,
  ``Three-Dimensional O(N) Theories at Large Distances,''
  Phys.\ Rev.\  D {\bf 24}, 2169 (1981).
}

\lref\AltlandZZ{
  A.~Altland and M.~R.~Zirnbauer,
  ``Nonstandard symmetry classes in mesoscopic normal-superconducting hybrid
  structures,''
  Phys.\ Rev.\  B {\bf 55}, 1142 (1997).
}

\lref\Ludwigtenfold{
Andreas P. Schnyder and Shinsei Ryu  and Akira  Furusaki and Andreas W. W. Ludwig, 
``Classification of topological insulators and superconductors in three spatial dimensions,"
  Phys.\ Rev.\  B {\bf 78}, 195125 (2008).
 }

\lref\Kitaevtenfold{
Alexei Kitaev,
``Periodic table for topological insulators and superconductors,"
arXiv:0901.2686.
}

\lref\WilczekMV{
  F.~Wilczek,
  ``Two Applications of Axion Electrodynamics,"
  Phys.\ Rev.\ Lett.\  {\bf 58}, 1799 (1987).
}

\lref\adler{
Stephen L. Adler,
``"Axial-Vector Vertex in Spinor Electrodynamics,"
Phys. Rev. {\bf 177}, 2426 (1969).
}

\lref\belljackiw{
J. S. Bell and R. Jackiw,
``A PCAC puzzle: $\eta^0 \rightarrow \gamma \gamma$ in the $\sigma$-model"
 Nuovo Cim.A60:47-61,1969.
}

\lref\fujikawa{
Kazuo Fujikawa,
``Path Integral for Gauge Theories with Fermions,"
Phys. Rev. D {\bf 21}, 2848, (1980).
}

\lref\WittenFP{
  E.~Witten,
  ``An SU(2) anomaly,''
  Phys.\ Lett.\  B {\bf 117}, 324 (1982).
}

\lref\wittenjones{
  E.~Witten,
  ``Quantum field theory and the Jones polynomial,''
  Commun.\ Math.\ Phys.\  {\bf 121}, 351 (1989).
}

\lref\NielsenRZ{
  H.~B.~Nielsen and M.~Ninomiya,
  ``Absence Of Neutrinos On A Lattice. 1. Proof By Homotopy Theory,''
  Nucl.\ Phys.\  B {\bf 185}, 20 (1981)
  [Erratum-ibid.\  B {\bf 195}, 541 (1982)].
}

\lref\NielsenXU{
  H.~B.~Nielsen and M.~Ninomiya,
  ``Absence Of Neutrinos On A Lattice. 2. Intuitive Topological Proof,''
  Nucl.\ Phys.\  B {\bf 193}, 173 (1981).
}

\lref\NepomechieKK{
  R.~I.~Nepomechie,
  ``Evaluating Fermion Determinants Through The Chiral Anomaly,''
  Annals Phys.\  {\bf 158}, 67 (1984).
}

\lref\AitchisonPP{
  I.~J.~R.~Aitchison and C.~M.~Fraser,
  ``Derivative Expansions Of Fermion Determinants: Anomaly Induced Vertices,
  Goldstone-Wilczek Currents And Skyrme Terms,''
  Phys.\ Rev.\  D {\bf 31}, 2605 (1985).
}

\lref\roy{
Rahul Roy,
``Topological phases and the quantum spin Hall effect in three dimensions,"
Phys. Rev. B {\bf 79}, 195322 (2009).
}

\lref\moorebalents{
J.~E.~Moore and L.~Balents,
``Topological invariants of time-reversal-invariant band structures,"
Phys. Rev. B 75, {\bf 121306} (2007).
}

\lref\fukanemele{
Liang Fu, C.~L.~Kane, and E.~J.~Mele,
``Topological Insulators in Three Dimensions,"
Phys. Rev. Lett. {\bf 98}, 106803 (2007).
}

\lref\emv{
Andrew~M.~Essin, Joel~E.~Moore, and David~Vanderbilt
``Magnetoelectric Polarizability and Axion Electrodynamics in Crystalline Insulators,"
Phys. Rev. Lett. {\bf 102}, 146805 (2009).
}

\lref\FradkinPQ{
  E.~H.~Fradkin, E.~Dagotto and D.~Boyanovsky,
  ``Physical Realization of the Parity Anomaly in Condensed Matter Physics,"
  Phys.\ Rev.\ Lett.\  {\bf 57}, 2967 (1986)
  [Erratum-ibid.\  {\bf 58}, 961 (1987)].
}

\lref\tsemacone{
Wang-Kong~Tse and A.~H.~MacDonald
``Magneto-optical and Magneto-electric Effects of Topological Insulators in Quantizing Magnetic Fields,"
Phys. Rev. B {\bf 82}, 161104 (2010).
}

\lref\tsemactwo{
Wang-Kong~Tse and A.~H.~MacDonald
``Magneto-Optical Faraday and Kerr Effects in Topological Insulator Films and in Other Layered Quantized Hall Systems,"
Phys. Rev. B {\bf 84}, 205327 (2011).
}

\lref\rahullevitov{
Rahul Nandkishore and Leonid Levitov,
``Polar Kerr Effect and Time Reversal Symmetry Breaking in Bilayer Graphene,"
Phys. Rev. Lett. {\bf 107}, 097402 (2011).
}

\lref\lang{
R.~Lang, A.~Winter, H.~Pascher, H.~Krenn, X.~Liu, and J.~K.~Furdyna ,
``Polar Kerr effect studies of $Ga_{1?x}Mn_xAs$ epitaxial films"
Phys. Rev. B {\bf 72}, 024430 (2005).
}

\lref\PreskillFR{
  J.~Preskill,
  ``Gauge anomalies in an effective field theory,''
  Annals Phys.\  {\bf 210}, 323 (1991).
}

\lref\DHokerPH{
  E.~D'Hoker and E.~Farhi,
Nucl.\ Phys.\ B {\bf 248}, 59 (1984)..
}

\lref\NiemiRQ{
  A.~J.~Niemi and G.~W.~Semenoff,
  ``Axial Anomaly Induced Fermion Fractionization and Effective Gauge Theory Actions in Odd Dimensional Space-Times,''
Phys.\ Rev.\ Lett.\  {\bf 51}, 2077 (1983).
}

\lref\RaoBA{
  S.~Rao and R.~Yahalom,
Phys.\ Lett.\ B {\bf 172}, 227 (1986).
}

\lref\WessYU{
  J.~Wess and B.~Zumino,
  ``Consequences of anomalous Ward identities,''
Phys.\ Lett.\ B {\bf 37}, 95 (1971).
}

\lref\ChandrasekharanAG{
  S.~Chandrasekharan,
  ``Anomaly cancellation in (2+1)-dimensions in the presence of a domain wall mass,''
Phys.\ Rev.\ D {\bf 49}, 1980 (1994).
[hep-th/9311050].
}

\lref\WittenEB{
  E.~Witten,
  ``Superconducting Strings,''
Nucl.\ Phys.\ B {\bf 249}, 557 (1985).
}

\lref\NaculichCI{
  S.~G.~Naculich,
  ``Axionic Strings: Covariant Anomalies And Bosonization Of Chiral Zero Modes,''
Nucl.\ Phys.\ B {\bf 296}, 837 (1988).
}

\lref\PolychronakosME{
  A.~P.~Polychronakos,
  ``Topological Mass Quantization And Parity Violation In (2+1)-dimensional QED,''
Nucl.\ Phys.\ B {\bf 281}, 241 (1987).
}

\lref\ClossetVP{
  C.~Closset, T.~T.~Dumitrescu, G.~Festuccia, Z.~Komargodski and N.~Seiberg,
  ``Comments on Chern-Simons Contact Terms in Three Dimensions,''
JHEP {\bf 1209}, 091 (2012).
[arXiv:1206.5218 [hep-th]].
}

\lref\WittenYA{
  E.~Witten,
  ``SL(2,Z) action on three-dimensional conformal field theories with Abelian symmetry,''
In *Shifman, M. (ed.) et al.: From fields to strings, vol. 2* 1173-1200.
[hep-th/0307041].
}

\lref\WenMW{
  X.~G.~Wen,
  ``Gapless Boundary Excitations In The Quantum Hall States And In The Chiral Spin States,''
Phys.\ Rev.\ B {\bf 43}, 11025 (1991).
}

\lref\StoneIW{
  M.~Stone,
  ``Edge Waves In The Quantum Hall Effect,''
Annals Phys.\  {\bf 207}, 38 (1991).
}

\lref\ElitzurNR{
  S.~Elitzur, G.~W.~Moore, A.~Schwimmer and N.~Seiberg,
  ``Remarks on the Canonical Quantization of the Chern-Simons-Witten Theory,''
Nucl.\ Phys.\ B {\bf 326}, 108 (1989).
}

\lref\WittenHF{
  E.~Witten,
  ``Quantum Field Theory and the Jones Polynomial,''
Commun.\ Math.\ Phys.\  {\bf 121}, 351 (1989).
}

\lref\WenEZ{
  X.~G.~Wen and Q.~Niu,
  ``Ground State Degeneracy Of The FQH States In Presence Of Random Potential And On High Genus Riemann Surfaces.''
}

\lref\WenIV{
  X.~G.~Wen,
  ``Topological Order In Rigid States,''
Int.\ J.\ Mod.\ Phys.\ B {\bf 4}, 239 (1990).
}

\lref\KaplanBT{
  D.~B.~Kaplan,
  ``A Method for simulating chiral fermions on the lattice,''
Phys.\ Lett.\ B {\bf 288}, 342 (1992).
[hep-lat/9206013].
}

\lref\ItzyksonRH{
  C.~Itzykson and J.~B.~Zuber,
  ``Quantum Field Theory,''
New York, Usa: Mcgraw-hill (1980) 705 P.(International Series In Pure and Applied Physics).
}

\lref\stonegoldbart{
  M.~Stone and P.~Goldbart,
  ``Mathematics for Physics: A Guided Tour for Graduate Students,''
Cambridge University Press; 1 edition (August 10, 2009).
}

\lref\BoyanovskyDT{
  D.~Boyanovsky, E.~Dagotto and E.~H.~Fradkin,
  ``Anomalous Currents Induced Charge And Bound States On A Domain Wall Of A Semiconductor,''
Nucl.\ Phys.\ B {\bf 285}, 340 (1987).
}

\lref\PeskinEV{
  M.~E.~Peskin and D.~V.~Schroeder,
Reading, USA: Addison-Wesley (1995) 842 p.
}

\lref\HaldaneZZA{
  F.~D.~M.~Haldane,
  ``Model for a Quantum Hall Effect without Landau Levels: Condensed-Matter Realization of the 'Parity Anomaly',''
Phys.\ Rev.\ Lett.\  {\bf 61}, 2015 (1988).
}

\lref\DeserVY{
  S.~Deser, R.~Jackiw and S.~Templeton,
  ``Three-Dimensional Massive Gauge Theories,''
Phys.\ Rev.\ Lett.\  {\bf 48}, 975 (1982).
}

\Title
{\vbox{
\baselineskip12pt
}}
{\vbox{
\baselineskip22pt 
{\centerline{Topological Insulators Avoid the Parity Anomaly}
\centerline{ } 
}}}

\centerline{Michael Mulligan$^{1}$ and F. J. Burnell$^{2, 3}$}
\bigskip
{\it \centerline{$^1$Station Q, Microsoft Research, Santa Barbara, CA 93106-6105}}
{\it \centerline{$^2$Theoretical Physics, Oxford University, 1 Keble Road, Oxford, OX1 3NP, United Kingdom }}
{\it \centerline{$^3$All Souls College, Oxford, OX1 4AL, United Kingdom}}

\medskip

\centerline{email: micmulli@microsoft.com}

\bigskip

The surface of a 3+1d topological insulator hosts an odd number of gapless Dirac fermions when charge conjugation and time-reversal symmetries are preserved.  
Viewed as a purely 2+1d system, this surface theory would necessarily explicitly break parity and time-reversal when coupled to a fluctuating gauge field.
Here we explain why such a state can exist on the boundary of a 3+1d system without breaking these symmetries, even if the number of boundary components is odd. 
This is accomplished from two complementary perspectives: topological quantization conditions and regularization.
We first discuss the conditions under which (continuous) large gauge transformations may exist when the theory lives on a boundary of a higher-dimensional spacetime. 
Next, we show how the higher-dimensional bulk theory is essential in providing a parity-invariant regularization of the theory living on the lower-dimensional boundary or defect.

\Date{January 2013}

\newsec{Introduction}

It is well known that a 2+1d theory consisting of an odd number of gapless Dirac fermions interacting with a fluctuating gauge field must break parity ($P$) and time-reversal ($T$) symmetries \refs{\RedlichKN, \RedlichDV, \AlvarezGaumeIG}.
It is also well-established that there exist topologically non-trivial band structures for fermions on a lattice in three spatial dimensions (i.e., 3+1d topological insulators) whose surfaces harbor an odd number of gapless Dirac fermions \refs{\fukanemele, \moorebalents,\roy}.  
This raises the question: Must topological insulators break $P$ and $T$ on their boundaries when coupled to a fluctuating gauge field?
The purpose of this paper is to explain why this does not occur.

Strong topological insulators in 3+1d (3DTI) are distinguished by the presence of gapless surface states that are stable to all $P$ and $T$ invariant perturbations (with respect to the boundary theory) that conserve electric charge \fukanetopins.  
Provided that the chemical potential is fine-tuned to the Dirac point, these gapless surface modes can be described at low energies by an odd number of 2-component Dirac fermions, which are charged under the electromagnetic $U(1)_{{\rm EM}}$ gauge field.
\foot{To be specific, the $U(1)_{{\rm EM}}$ gauge field is not confined to the 2+1d surface; it is free to explore the 3+1d bulk. 
Integrating over this extra direction results in a tree-level propagator for the gauge field that is softer in the infrared (IR) in that it diverges linearly as opposed to quadratically at small momentum.}
In a purely 2+1d theory of this type (i.e., QED3), gauge invariance (under both large and small gauge transformations) is preserved if and only if
\eqn\constraint{{N_f \over 2} + k \in {\bf Z},
}
where $N_f$ is the number of flavors of 2-component Dirac fermions and $k$ is the level of the Chern-Simons (CS) term for the gauge field.
\foot{This statement is slightly imprecise. 
Invariance of the theory under large gauge transformations requires a choice of auxiliary 4-manifold to ensure the CS term is well defined in general \refs{\WittenYA, \ClossetVP}.
While this more precise definition is suggestive of our eventual conclusion, we will not make use of it further.}
When $N_f$ is odd, a non-zero half-integral level CS term must supplement the effective action.  

This is known as the parity anomaly. 
It is the gauge-invariant regularization of the theory that results in the addition of the half-integral level CS term to the effective action when $N_f$ is odd.
This CS term explicitly breaks $P$ and $T$.
Physically, the anomaly means that parity and time-reversal invariance are explicitly broken when a theory with an odd number of 2+1d Dirac fermions is coupled to a fluctuating gauge field.  

Anomaly considerations establish a relationship between certain topologically ordered phases of matter and the gapless modes living on their boundaries \refs{\ElitzurNR, \WenMW, \StoneIW, \RyuMooreLudwig}.
The quantum Hall effect provides, perhaps, the most famous example of how anomaly considerations can be used to better understand these gapless boundary modes
in a model independent way \refs{\WenMW, \StoneIW}.
In this example, the $U(1)_{{\rm EM}}$ charge conservation symmetry is gauged by the electromagnetic field, whose (possibly fractionally) quantized low energy Hall response implies an effective bulk description by a CS theory.  
Equivalently, the non-dissipative Hall current requires charge-carrying chiral edge modes.
The potentially anomalous $U(1)_{{\rm EM}}$ gauge symmetry provides the link between these two descriptions.
In the presence of a boundary, neither the bulk CS theory nor the boundary chiral theory is individually gauge invariant;
however, their anomalous variations cancel one another so that the underlying $U(1)_{{\rm EM}}$ symmetry is maintained in the system as a whole.  Thus in the presence of a boundary, the bulk theory cannot exist without the gapless boundary modes, and vice versa.   

This effect is known as anomaly inflow \CallanSA.
A classical anomaly of the bulk effective action is cancelled by a {\it local} quantum mechanical anomaly of the boundary theory.
Intuitively, the cancellation occurs because charge flows out from the bulk and along the boundary of the system at a rate determined by the applied external field.  Because this relationship readily extends to interacting systems, it is a powerful demonstration of the robustness of the gapless boundary modes \RyuMooreLudwig.

Anomaly inflow does not, however,  define the relationship between bulk and boundary theories for 3+1d topological insulators.  
This is because local violation of charge or momentum conservation can only occur when the spatial dimension is odd \refs{\adler, \belljackiw, \fujikawa, \AlvarezGaumeIG}, as is the case for the 1+1d boundary of a 2+1d quantum Hall system.
Instead, possible anomalies relevant to theories in even spatial dimension necessarily involve so-called large gauge transformations.
This is the case for the parity anomaly constraint \constraint, which is potentially relevant to the surface of a 3+1d topological insulator.
\foot{See \refs{\FradkinPQ, \BoyanovskyDT} for an earlier discussion on a closely related system.}  

A (topological) band insulator is necessarily a system that can be realized on the lattice so it is worth reviewing the conventional wisdom regarding anomalies in lattice systems.
Because the lattice itself provides a gauge-invariant regularization, any system that can be realized on the lattice cannot be anomalous.
That this is true follows immediately from fermion doubling \fermiondoubling.
For example, a purely 2+1d lattice system of fermions with relativistic dispersion necessarily contains an even number of (low energy) Dirac fermions in the absence of $P$ and $T$ breaking.
The parity anomaly constraint \constraint\ is satisfied by the low energy effective theory for all deformations preserving the $U(1)$ symmetry -- even those that break $P$ or $T$ \HaldaneZZA.  
The striking feature of topological insulators is that the gapless surface modes do not exhibit fermion doubling because they live on the boundary of a higher-dimensional system.

Thus, we must ask whether \constraint\ is obeyed on each boundary or, if not, how this is consistent with an overall gauge-invariant theory.
Because of the $Z_2$ nature of the parity anomaly, a topological insulator with two (separate) boundaries automatically satisfies \constraint.
However, there is no fundamental reason why a topological insulator must have an even number of boundaries; 
for example, a solid sphere or torus of 3+1 d topological insulator would have a single boundary, potentially violating \constraint.

In this work, we explain why \constraint\ need not be satisfied by the low energy theory describing surface Dirac fermions interacting with a bulk gauge field.
Specifically, we shall explain how the higher-dimensional bulk theory from which the surface modes descend eliminates the potential anomaly of the surface. 
We emphasize that this conclusion is quite different from what occurs in the quantum Hall case \WenMW\ and other examples studied by \RyuMooreLudwig, where the potential anomaly of the surface is {\it cancelled} by a comparable anomaly in the bulk. 

In fact, we find that ${N_f \over 2} + k$ is half-integral at each boundary surface of a 3+1d topological insulator only in the limit that the bulk gap $m_0$ is infinitely large compared to any T-breaking perturbations on the boundary.
The corrections appearing at finite $m_0$ imply that the CS level  $k$ {\it need not be quantized} at half-integral (or indeed at any rational) values, even in a non-interacting system.

The conclusion that the surface states do not exhibit the parity anomaly --in the sense that they do not obey \constraint\ --  is of  clear importance for the low energy properties of a topological insulator: a non-zero bare CS term breaks $P$ and $T$, and would therefore drastically affect the low energy physics.  
While these surface properties are well established (or at least well believed) theoretically for the case of topological band insulators which need not be coupled to a fluctuating gauge field in order to be defined, our analysis is equally applicable to the case of more exotic fractional topological insulators, such as those described in \refs{\MaciejkoTX, \SwingleEtAl, \LevBur, \MacjeckoLong}, in which the presence of a fluctuating ``internal" gauge field is inevitable.
Thus, our result is important in clarifying why $P$ and $T$ invariant gapless boundary modes exist in these systems -- as well as in understanding their topological order.  

We would like to point out that in the context of topological insulators, a different definition of the parity anomaly is sometimes used.  It is sometimes said that the surfaces modes exhibit the parity anomaly because the parity-violating current equation,
\eqn\paritycurrent{\langle J_\mu \rangle = {1 \over 4 \pi} \epsilon_{\mu \nu \rho} \partial^\nu A^\rho,
}
is satisfied on any boundary perturbed by a $P$ and $T$ odd interaction.
For a topological insulator boundary, \paritycurrent\ does {\it not} mean that there exists a zero magnetic field Hall conductance (in contrast to a purely 2+1d system); rather, a Hall effect occurs only after a time-reversal breaking perturbation has been applied to the surface, as we shall explain. 
(\paritycurrent\ follows directly from the effective action calculated in \S4.)
In the present work, when we say that the surface modes of a 3+1d topological insulator do {\it not} exhibit the parity anomaly, we mean that they do not satisfy the constraint \constraint\ requiring integral $N_f/2 + k$ (on each boundary component).
It is important to note this difference in terminology in order to avoid possible confusion.
\foot{We thank E. Fradkin for discussions on this point.}

The remainder of the paper is organized as follows.
We begin in \S2 with a brief introduction to the specific model we wish to study.
We then turn to the explanation for why non-integral ${N_f \over 2} + k$ is consistent with a gauge-invariant theory.
To do so, we make use of two complementary perspectives: topological quantization conditions in \S3 and perturbative regularization in \S4.
We summarize and conclude in \S5.

The paper contains five (count them!) appendices summarizing issues that are related, but not essential to the above line of argument, although they may be of some interest.
Appendix A contains a review of domain wall fermions and their relation to continuum models of topological insulators.
In Appendix B, we discuss anomaly inflow intuition for line and domain wall defects in three spatial dimensions.
In Appendix C, we recall how flux insertion arguments can be used to define a strong topological insulator in the presence of disorder or other interactions. 
In Appendix D, we repeat the perturbative analysis of \S4 in the technically simpler, but conceptually equivalent case of 1+1d allowing direct contact with the work of Goldstone and Wilczek \GoldstoneKK.
In Appendix E, we elaborate in detail upon the leading divergence structure arising from the interaction between bulk and boundary modes in our toy model of a topological insulator, thereby confirming the conclusions of \S4.

\newsec{Domain Wall Fermions, the $\theta$-term, and Topological Insulators}

There are four $Z_2$ topological invariants that characterize the bulk band structure of (non-interacting) fermionic insulators in 3+1d \refs{\fukanemele, \moorebalents,\roy}.  
These distinguish between three classes of time-reversal invariant, charge-conserving band insulators: a so-called trivial insulator (with no protected low-energy surface modes), a ``weak" topological insulator (which has gapless surface states that can be gapped without breaking time-reversal symmetry, but are nonetheless robust to disorder \SternWeak), and a ``strong" topological insulator (STI), whose gapless surface states cannot be eliminated by any time-reversal invariant perturbation.
\foot{It is well established that this is the case for non-interacting systems; however, the arguments of \FuKanePump, applied in 3D as explained in \LevBur, strongly suggest that this remains true in the presence of interactions.}
In this paper, we focus on the STI.  

At low energies, a STI in 3+1d can be described by a continuum theory of a single, massive 4-component Dirac fermion \refs{\fukane, \qhz}.  
In this continuum formulation, the STI is distinguished from its trivial counterpart by the sign of the Dirac fermion mass $m$.
The existence of two distinct insulators distinguished by the sign of $m$ is the continuum version of the notion of topological band structure:
these two insulators cannot be adiabatically connected without either closing the bulk gap or choosing a connecting path in parameter space that breaks $P$ and $T$ (e.g., by interpolating over complex fermion masses).

To exhibit the difference, we consider the action,
\eqn\fouraction{S = \int d^4 x \Big( \bar{\psi} i \gamma^\mu (\partial_\mu - i e A_\mu) \psi + m(x) \bar{\psi} \psi \Big),
}
where $\psi$ is a 4-component spinor that describes the massive bulk fermion with spatially dependent mass $m(x)$ coupled to the $U(1)$ gauge field $A_\mu$
which may represent the electromagnetic field.
The spatially-varying mass allows us to study the interface between a topologically non-trivial and a topologically trivial insulator, where the parity anomaly constraint \constraint\ could potentially be applied.  
(We have not included kinetic terms for the gauge field, since their specific form does not affect our results.) 
In \fouraction, $\bar{\psi} = \psi^{\dagger} \gamma^0$, $\{\gamma^\mu, \gamma^\nu\} = 2 \eta^{\mu \nu}$ for $\mu, \nu = 0, 1,2,3$ where $\eta^{\mu \nu} = {\rm diag}(1, - 1, -1,-1)$. 
Further, we define the matrix $\gamma_5 = i \gamma^0 \gamma^1 \gamma^2 \gamma^3$ 
which anti-commutes with all $\gamma_\mu$.

Let us begin by reviewing the claim that the sign of the fermion mass distinguishes between two distinct bulk phases of matter. 
If $m(x)$ is constant, the two states are distinguished by the presence or absence of a topological $\theta$-term in their low-energy effective action \refs{\qhz, \WilczekMV, \emv}.
We can see this by considering the effect of chiral rotations $\psi \rightarrow \exp(i \alpha \gamma_5) \psi$ on the effective action \fouraction .  
These rotate the fermion mass according to
\eqn\complexmass{m \bar{\psi} \psi \rightarrow m \bar{\psi} e^{i 2 \alpha \gamma_5} \psi = m \cos(2 \alpha)\bar{\psi} \psi + i m \sin(2 \alpha) \bar{\psi} \gamma_5 \psi
}
rendering it complex unless $\alpha$ is an integer multiple of $\pi/2$.
Importantly, chiral rotations also contribute an anomalous term to the action from the path integral measure:  
\eqn\topterm{S_{\theta} = \left( 2 \alpha \right) \int d^4x {e^2 \over 32 \pi^2} \epsilon_{\mu \nu \rho \sigma} F_{\mu \nu} F_{\rho \sigma}.
}
By the chiral rotation $2 \alpha = \pi$ -- which is nothing more than a change of path integration variables -- we may change the sign of the fermion mass at the expense of creating a topological $\theta$-term \topterm\ with coefficient $\theta = \pi$.  
Thus, the effective continuum actions for the topologically non-trivial and trivial insulators with constant masses everywhere in space, but with opposite sign, differ precisely by the topological term \topterm\ with $\theta = \pi$.
 
Now consider the scenario in which there is a single domain wall separating a region of STI ($x_3 >0$) from a region of the vacuum or trivial insulator ($x_3<0$). 
As we review in Appendix A,
if we take
\eqn\asymptotic{\lim_{x_3 \rightarrow \pm \infty} m(x_3) = \pm m_0, \ \ m_0 >0}
with $m(x_3)$ passing through zero exactly once, at $x_3=0$, 
there is a single massless 2+1d Dirac fermion localized near $x_3=0$, where $m(x)$ changes sign  \JackiwFN .    

Based on the result \topterm , we might expect a $\theta$-term with a spatially varying coefficient taking the value, say, $\theta =0$ for $x_3$ large and negative, and $\theta = \pi$ for $x_3$ large and positive.
A bulk $\theta$-term integrates by parts to a boundary CS term at level $k = \theta/2\pi$.
This suggests that the low energy action might obey \constraint with the half-integer CS level, obtained from a bulk $\theta$-term, compensating for the odd number of domain wall fermions.
If true, this would mean that $P$ and $T$ are explicitly broken at the surface of a topological insulator in the absence of any such symmetry-violating perturbations.
In particular, a magnetic perturbation that opens up a gap in the surface fermion spectrum would imply an {\it integral} Hall effect, along with the consequent change of the Kerr and Faraday angles for light passing through a single surface \refs{\qhz, \tsemacone, \tsemactwo}.

In the remainder of the paper, we show that this is not the case: we should think of the domain wall as associated {\it either} with the presence of an odd number of gapless 2+1d Dirac fermions, {\it or} (if these are gapped) with a half-integral surface Hall conductivity which may be understood as arising from a spatially varying $\theta$-term.    
This conclusion is equally true for theories of fermions coupled to other, possibly non-Abelian, gauge fields, as is relevant for describing strongly interacting fractional topological insulators.  
In other words, the surface of the topological insulator fails to obey \constraint; ${N_f \over 2} + k$ is half-integral (in the limit $m_0 \rightarrow \infty$).

\newsec{Large Gauge Transformations, the Parity Anomaly, and Topological Insulators}

In order to understand the applicability of constraint \constraint\ to the surface modes of a topological insulator, we first review its derivation from the perspective of topological quantization conditions in 2+1d.
The generalization of this logic to topological insulators is then immediate.

\subsec{The Fermion Determinant}

Recall that there are two types of possible gauge anomalies of an effective action: local or global.
(For general discussions, see \refs{\PeskinEV, \AlvarezGaumeIG, \PreskillFR}.)
The distinction arises from the class of the particular gauge transformation under which the action fails to be invariant.
If a local anomaly is present, the action fails to be invariant under any gauge transformation that is continuously deformable to the constant map.
The current associated with a locally anomalous symmetry fails to be conserved. 

In contrast, the parity anomaly is an example of a global anomaly: an anomaly associated with so-called ``large" gauge transformations.
By a large gauge transformation, we mean a gauge transformation that is not continuously deformable to the constant map. 
For example, two maps with distinct winding number from the circle to itself cannot be continuously deformed into one another.

Invariance of the effective action of QED3 under large gauge transformations requires that ${N_f \over 2} + k$ be integral.
To see this, consider first the fermionic contribution to the gauge field effective action,
\eqn\fermdet{e^{ i S_F(A) } = \int [d \psi] [d\bar{\psi}] \exp\Big(\int \bar{\psi} (iD_3^{(A)} - m_0) \psi\Big)  = \Big(\det(D_3^{(A)} - m_0)\Big)^{N_f/2} ,
}
where $D_3^{(A)} = \tilde{\gamma}^a (\partial_a - i e A_a)$ is the 2+1d Dirac operator for a given gauge field configuration $A$.
The fermion determinant is essentially a product over the energies of the fermionic states.  
These energies are negative for states that lie below the chemical potential, and positive for states above it.  
(In this paper, the chemical potential is fine-tuned to zero so that charge-conjugation symmetry in maintained.)
The square root instructs us to only include, say, the filled negative energy states in the product \fermdet.

Therefore, an anomalous transformation of the fermion determinant (or anomaly, for short) occurs when an odd number of fermions are ``pumped" from immediately below the Fermi surface to states immediately above it under a large gauge transformation.
(Gauge transformations deformable to the identity can have no such effect.)
In such a situation, the fermion determinant changes sign.
This renders the partition function, which is a sum over all such sectors, ill defined if there is no compensating bare CS term.

As a concrete example for how this works, consider a Dirac fermion on a spatial torus $S^1\times S^1$.
We will also assume that all gauge fields tend to constant values and so are pure gauge as $t\rightarrow \pm \infty$, effectively imposing periodic boundary conditions in time on all physical observables.
Suppose the fermions are given anti-periodic boundary conditions along the two cycles of length $L_1$ and $L_2$.
Then the allowed fermion momenta (in the absence of any external gauge field) are
$
k_1 = { 2 \pi \over L_1 }(n_1 + {1 \over 2} ), \   k_2 =  {2 \pi \over L_2} (n_2 + {1 \over 2} )
$
with associated band energies $E = \pm \sqrt{k_1^2 + k_2^2}$.
Consider the Dirac cone sitting at the time-reversal invariant momentum point $(k_1, k_2) = (0,0)$.
The allowed momentum states are distributed symmetrically about the Dirac point $(0,0)$ with the lowest energy states at $(k_1, k_2 ) =  (\pm {\pi \over L_1}, \pm {\pi \over L_2})$.  

Large gauge transformations in this system correspond to inserting magnetic flux quanta $(l_1 ={1 \over \Phi_0} \oint d x_1 A_1, l_2={1 \over \Phi_0} \oint d x_2 A_2)$ through the two non-contractible curves of the torus where the magnetic flux quantum is $\Phi_0 = 2\pi/e$.  
The initial choice of  $l_1$ and  $l_2$ has no effect on the physical spectrum or the allowed momenta: they are all gauge equivalent.
When $l_1$ or $l_2$ is adiabatically changed  from one value to another, however, we must pass through intermediate flux configurations that cannot simply be gauged away.  
At a generic intermediate point in the variation, there is an effect on the boundary conditions for the fermions, which now feel a Berry phase due to the magnetic flux as they encircle the torus.  
If the flux through the two holes of the torus is $(\Phi_1, \Phi_2)$, the allowed momenta
$
k_1 = { 2 \pi \over L_1 }(n_1 + {1 \over 2} + { \Phi_1 \over 2 \pi}), \  k_2 =  {2 \pi \over L_2} (n_2 + {1 \over 2} +{ \Phi_2 \over 2 \pi}).
$

To see the anomaly in action, let us track the fermion spectrum under a large gauge transformation in which we simultaneously increase the flux through the two non-contractible curves from $0$ to $2 \pi$.
It is necessary to insert non-trivial flux through both cycles; otherwise, there is no zero crossing.
As we increase the flux from $(0, 0)$ to $(\pi, \pi)$, the lowest energy fermion states at $(k_1, k_2 ) =  (\pm {\pi \over L_1}, \pm {\pi \over L_2})$ are shifted so that they sit at $(k_1, k_2 ) = ( {2 \pi \over L_1},  {2 \pi \over L_2}), (0,  {2 \pi \over L_2}), ( {2 \pi \over L_2},0),$ and $(0,0)$, as shown in Fig. 1a.   
If we now continue slowly inserting flux, the energies of the states at  ($ {2 \pi \over L_1},  {2 \pi \over L_2}), (0,  {2 \pi \over L_2}), ( {2 \pi \over L_2},0)$ will remain negative, since they are separated by a non-zero gap from the positive energy states in the band above.  
However, the energy of the state at $(0,0)$, which is exactly $0$, will change sign, so that this state arrives at its final momentum $( {\pi \over L_1}, {\pi \over L_2})$  with {\it positive} energy, having moved up to the conduction band (Fig. 1 b).
Because of the single zero crossing, the fermion determinant changes sign.
\bigskip
\bigskip
\centerline{
\epsfxsize 3truein\epsfbox{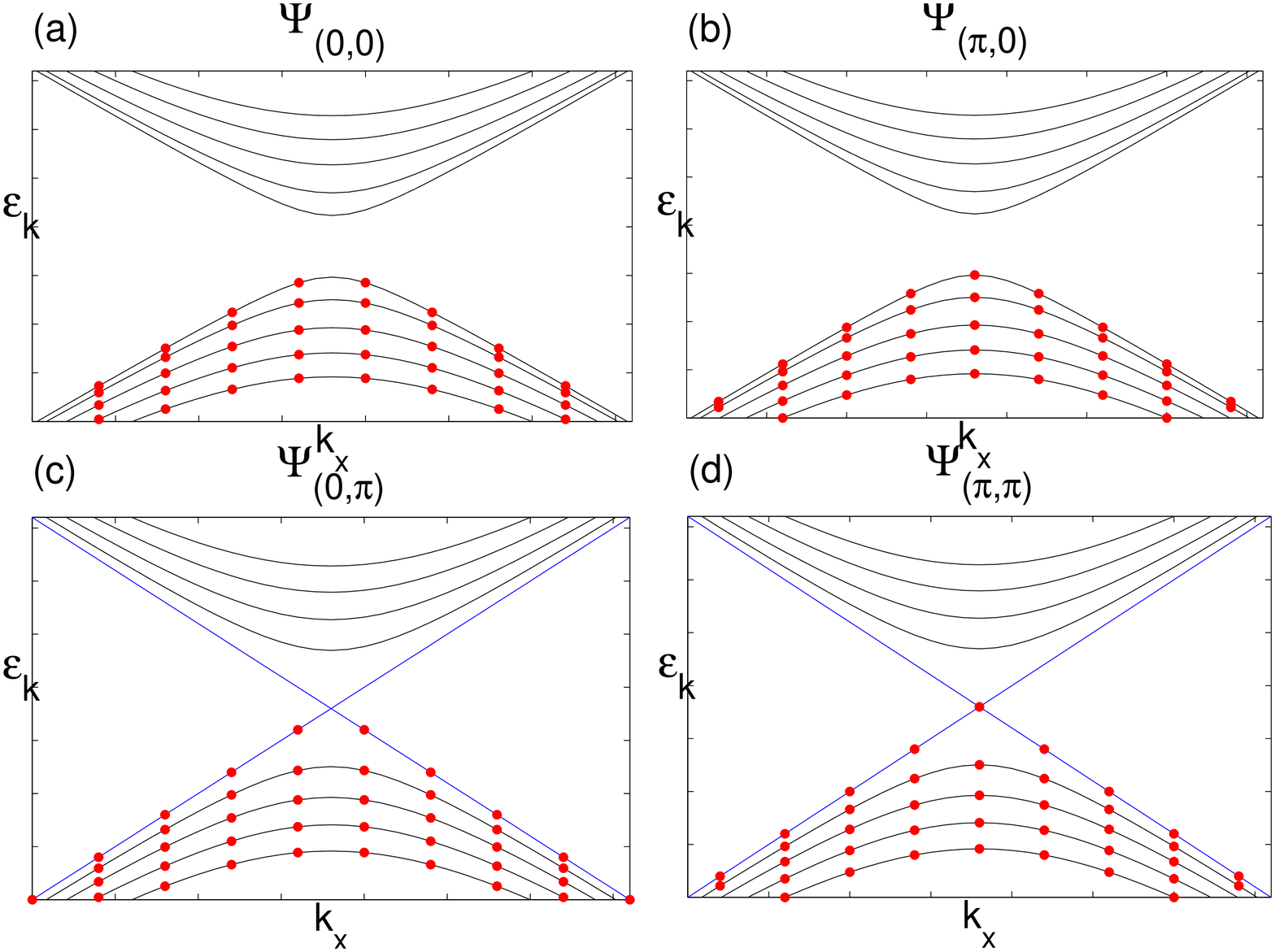}
\epsfxsize 3truein\epsfbox{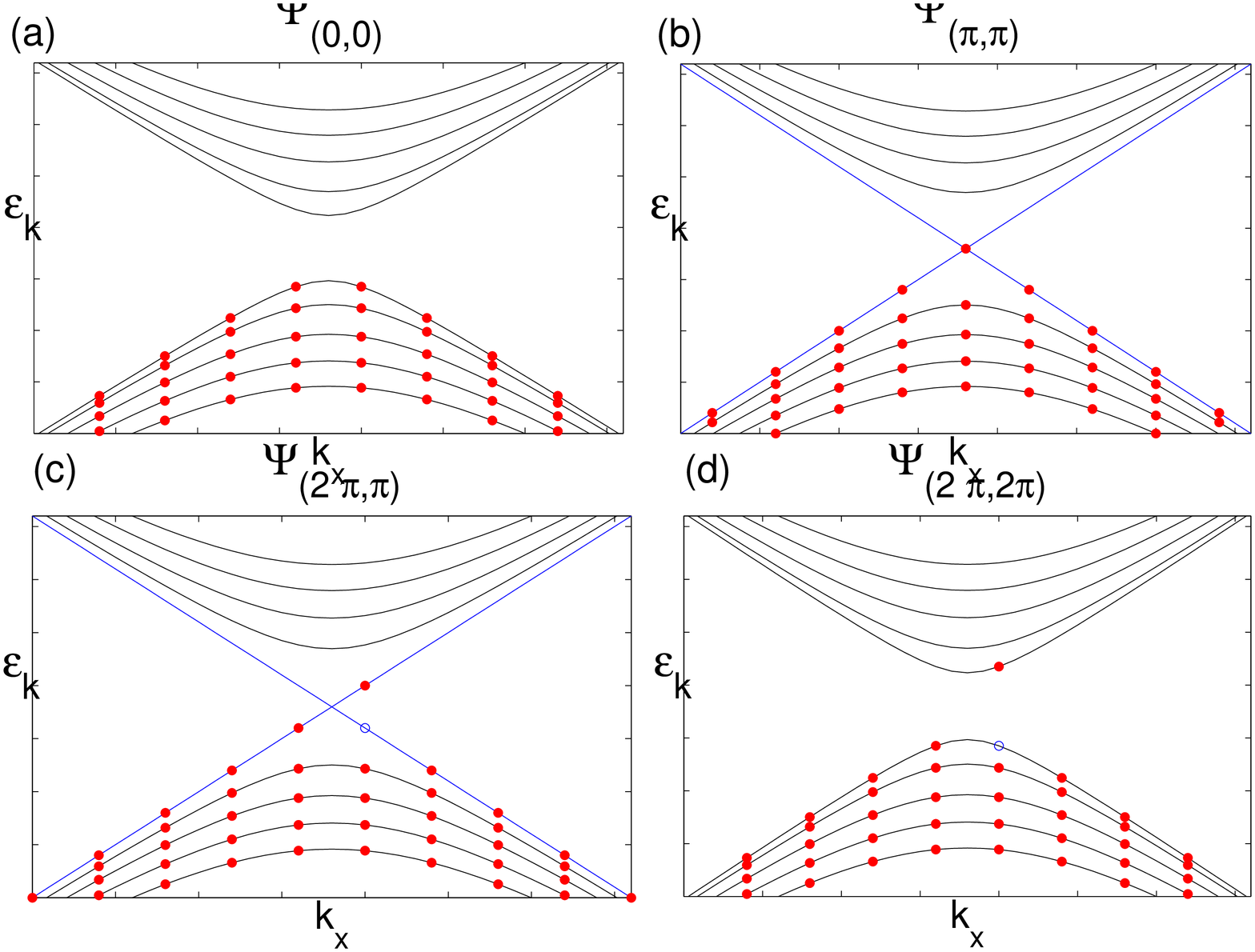}}
\centerline{\ninepoint
\baselineskip=2pt {\bf Fig. 1.} {Flux insertion and Kramers degeneracy for non-interacting fermions on the surface of the torus.}}
\bigskip

How do we know that a filled state from the lower band must move to the upper band during this process?
Let $|\Omega \rangle$ be the many-body ground state living on the spatial torus and call the operator that inserts $(\pi, \pi)$ flux through the two cycles of the torus $\Phi_{\pi, \pi}$.
Consider the two many-body states $\Phi_{\pi, \pi} |\Omega \rangle$ and  $T \Phi_{\pi, \pi} |\Omega \rangle$.
Kramers degeneracy requires that $\langle T \Phi_{\pi, \pi} \Omega |\Phi_{\pi, \pi}\Omega \rangle = 0$.
The reason is that the state transported to the tip of the Dirac cone after an insertion of $(\pi, \pi)$ flux is orthogonal to its time-reversed partner.
(For example, $T$ transforms the spin-up state at momentum $(0,0)$ to its spin-down partner at $(0,0)$.)  
We then apply $\Phi_{\pi, \pi}$ to each many-body state to obtain $\Phi_{\pi, \pi} \Phi_{\pi, \pi} |\Omega \rangle$ and $\Phi_{\pi, \pi} T \Phi_{\pi, \pi} |\Omega \rangle \sim |\Omega \rangle$.
The resulting many-body states remain orthogonal.
Since the many-body ground state with an integral multiple of $2\pi$ flux inserted through each cycle is non-degenerate, the resulting states can be orthogonal only if at least one of the states below  the Fermi level is now empty.
In other words, there has been a zero crossing.

More generally, the fact that there are always an odd number of zero crossings during the above flux insertion process follows from a 3+1d Atiyah-Singer index theorem \PolychronakosME.
Let $\tau$ parametrize the ``time" over which the flux insertion process occurs.
Using the auxiliary $\tau$ direction, we may construct the 3+1d Dirac operator $D_4^{(A)} = \gamma^\mu(\partial_\mu - i e A_\mu)$ for $\mu = 0, ..., 3$ with $A_3 = 0$ and
$\gamma^i = \sigma_3 \otimes \tilde{\gamma}^i$ for $i = 0, 1,2$ and $\gamma^3 = \sigma_1 \otimes 1_{2 \times 2}$.
The determinant of $D_3^{(A)}$ is obtained from the square root of the determinant of $D_4^{(A)}$. 
The number of zero crossings under an adiabatic flux insertion process is equal to the number of zero modes of the four-dimensional Dirac operator $D_4^{(A)}$ \WittenFP.
The index theorem ensures that the number of zero crossings equals $\Delta l_1 \cdot \Delta l_2$.
It is important that the change in flux number $\Delta l_i$ through each cycle is non-zero; otherwise, there would be no zero crossing.

\subsec{The Chern-Simons Term}

When there is an even number of Dirac cones ($N_f$ is even), there can be no sign change of the fermion determinant under any large gauge transformation. 
If $N_f$ is odd, however, we require a bare half-integral level CS term to compensate for the anomaly.

To see this, consider $U(1)$ CS theory at level $k$ on the spatial torus $S^1 \times S^1$,
 \eqn\Scs{ S_{CS} = {k \over 4 \pi} \int_{(S^1)^3} \epsilon_{\mu \nu \rho} A_\mu \partial_\nu A_\rho.
 }
The gauge coupling $e$ has been set to unity in this section.
We impose periodic boundary conditions $x_1 \equiv x_1 + L_1, x_2 \equiv x_2 +L_2$ on the gauge field and the condition that $A_\mu$ is pure gauge as $t \rightarrow \pm \infty$ implies that our spacetime is effectively $S^1 \times S^1 \times S^1$.
(The time direction can be thought of as the unit circle if we require $A_\mu(x,y, t \rightarrow - \infty) =A_\mu(x,y, t \rightarrow \infty)$  up to gauge transformations.)
Consider a field configuration with $l_1$ flux quanta passing through the non-contractible curve along $x_1$:
\eqn\monopole{ \int d x_1 A_1 = 2 \pi l_1,}
Let us now adiabatically insert $l_2$ flux quanta through the non-contractible curve in $x_2$.  
To do this we must generate an infinitesimal electric field $E_2(t) = {2 \pi l_2 \over L_y}, 0 \leq t \leq 1$ (for sufficiently large $L_2$).  
The change in the Chern-Simons term over the course of this flux insertion is
 \eqn\abtrans{\delta S_{CS} = { k \over 2 \pi} \int_0^1 dt \int d x_1 dx_2 A_1 E_2 (t)  =  2 \pi l_1 l_2 k .
}
$k$ plays the same role as ${N_f \over 2}$.
This shows that if $k$ is a half-integer, $S_{CS}$ changes by $\pi$ if we insert a single flux quantum through the both non-contractible curves along $x_1$ and $x_2$.   

Thus, both the Chern-Simons contribution to the partition function at $k = 1/2$ and the fermion determinant for $N_f$ odd change sign under large gauge transformations with $l_1 = l_2 =1$.   To avoid the parity anomaly, the overall sign of the gauge field partition function must be invariant under these large gauge transformations.  This is the origin of the condition \constraint.

The relationship between the parity anomaly of a single surface and the time-reversal protected surface states of the STI can be established without explicit knowledge of the surface band structure, as we discuss in Appendix C.  
This ensures that it remains valid if we tune the system away from vanishing chemical potential, or in the presence of disorder or strong interactions.   
Essentially, the statement that 
each surface, when viewed as an isolated system, violates the parity anomaly constraint \constraint\ is equivalent to the statement that a bulk system is a STI.  

The analysis above readily extends to more general (continuous) gauge groups.  
Indeed, the parity anomaly was originally discussed in the context of $SU(2)$ gauge theory where the analysis is mathematically somewhat simpler \refs{\RedlichKN, \RedlichDV, \AlvarezGaumeIG}.  
Let us briefly review this analysis as well since the language and notation will be useful later.

For a non-abelian gauge theory, say $SU(2)$, we may think of spacetime as being topologically the 3-sphere by imposing appropriate boundary conditions.
(The anomaly cares only about topology and not about whether the metric is Lorentzian or Euclidean.)
In general, we can think of gauge transformations as maps from spacetime into the gauge group.
Such maps are characterized by their degree or ``winding number."  
The possible winding numbers are determined by the group $\Pi_{3}(SU(2)) = {\bf Z}$ and so non-trivial large gauge transformations are elements of non-zero degree.
(The winding numbers of a U(1) gauge transformation were denoted by $l_1$ and $l_2$.)

For $SU(2)$, the parity anomaly arises because of the following two facts.  
First, if the $SU(2)$ gauge field is coupled to $N_f$ flavors of 2-component Dirac fermions, then under large gauge transformations of degree $n$, the fermion determinant transforms by $(-1)^{n N_f}$; or equivalently, the gauge field effective action shifts by $\pi n N_f$.  
Second, if there is a bare Chern-Simons term in the action for our gauge field, this term also shifts by $2 \pi n k$ under a transformation of degree $n$ \DeserVY.  
The combination of these observations again gives us \constraint.

\subsec{Large Gauge Transformations and Domain Wall Fermions}

The discussion above highlights the importance of the topology of spacetime in deriving \constraint.
The essential difference between a topological insulator and these purely 2+1d systems is that while the low energy fermions are localized to a boundary or defect in the fermion mass (assuming a large bulk gap), the gauge field is free to propagate throughout the bulk.
(See Appendix A for a review of this fermion localization.)
Thus, we must consider the role played by the bulk geometry. 

Let us gain intuition by first studying the case where the gauge group is $SU(2)$, where the geometries involved are simpler.
There are two general cases to consider.
First, take the topology of spacetime to be the 4-ball. 
This space has a {\it single} 3-sphere boundary.
If we were to forget about the interior of the ball, we could again classify gauge transformations by $\Pi_3(SU(2)) = {\bf Z}$.
It is the gauge transformations of non-zero degree that lead to a quantization condition on the CS level $k$.
What becomes of them when we fill in the 3-sphere to make  the 4-ball?
Only transformations of degree zero extend continuously into the 4-ball bulk.
This means that non-trivial (from the perspective of the boundary theory) large gauge transformations are not allowed in the theory as they do not continuously extend; therefore, \constraint\ need not be imposed.

The second situation occurs when the system has topology, say, $S^3 \times I$, where $I$ is the unit interval.
In this case, the boundary has {\it two} components, each living on one end of the interval.
A smooth extension is now allowed, however, a gauge transformation of degree different from zero acts in precisely the same way on each boundary.
Because the anomalous transformation associated with each $S^3$ boundary component transforms the path integral at most by a phase, the phases associated with each boundary, being equal and opposite (due to their opposite relative orientations), precisely cancel one another.
Because the cancellation is automatic, there is no non-trivial constraint to impose.

For a general four-dimensional geometry with some number of boundary components, the existence of large gauge transformations follows from a basic result of homotopy called the extension lemma.
\foot{This lemma is familiar from the construction of Wess-Zumino-Witten (WZW) terms.
Such a WZW term is constructed by extending a field over an auxiliary higher-dimensional space.}
A map admits a continuous extension if and only if its total degree (i.e., the sum of its degrees restricted to each boundary component) vanishes.
This condition ensures that large gauge transformations (if they are allowed with non-trivial degree) have no effect on the partition function of any such theory as any accumulated phases (e.g., \abtrans\ from a CS term) must mutually cancel, i.e., there is no anomaly.

The application of this logic to a theory with $U(1)$ gauge group is now straightforward.
In the previous section, we considered a 2+1d theory living on $S^1_{(t)} \times S^1_{(x)} \times S^1_{(y)}$.
There are two simple ways to form a four-dimensional geometry.
First, we may fill in the interior of one of the circles of the 3-torus so that the entire system only contains a single boundary.
Without any loss of generality, we may work in $A_t = 0$ gauge and consider the field configuration,
$A_x = 2\pi l_1/L_x$ and $A_y = 2\pi l_2 t/L_y$ for $0 \leq t \leq 1$.
This configuration is pure gauge (i.e., it can be written as $A_a = e^{- i f(x,y)} \partial_a e^{i f(x,y)}$) at times $t =0$ and $t = 1$.
It has non-zero winding around $S^1_{(x)}$ and interpolates in time from an $l_2 = 0$ to an $l_2 = 1$ winding around $S_{(y)}^1$.
If we choose to fill in $S_{(x)}^1$, then $A_x$ is no longer pure gauge if $l_1 \neq 0$ as it cannot be removed by a gauge transformation continuously extendable into the bulk.
This means that only gauge configurations of zero winding in $x$ should be included in the physical theory.
Similarly, filling in either the $S^1_{(t)}$ or $S^1_{(y)}$ is only compatible with $\Delta l_2 = 0$.
A quick way of drawing the above conclusion is that, in the absence of any sources, the Wilson loop around $S^1_{(x)}$ or the flux through $S^1_{(t)} \times S^1_{(y)}$ must vanish if any of the $S^1_{(a)}$ circles is the boundary of some disc.

Alternatively, we may consider the four-dimensional geometry $(S^1)^3 \times I$.
This is the geometry we implicitly used in the previous section when we discussed flux insertion arguments. 
Large gauge transformations are allowed, but they impose no constraint on the physical theory as their effects on each boundary mutually cancel.

In summary, QED3 must satisfy \constraint\ by having integral ${N_f \over 2} + k$, if it is to be invariant under large gauge transformations.
The gauge field coupled to the low energy degrees of freedom of a topological insulator, however, is not localized to the boundary; rather, it explores the entire bulk.
Therefore, we must determine whether or not non-trivial large gauge transformations of the boundary theory extend smoothly into the bulk.
If they do not, then the constraint imposed by their existence is lifted.
In situations for which they do extend, the number of boundaries is even (or at least the sum of the ``degrees" of the maps restricted to each boundary vanishes) and the transformation acts identically on each boundary,  so that \constraint\ is trivially satisfied. That is to say, there is no non-trivial constraint.

\newsec{Regularizing the Boundary Theory}

Regularization provides a complementary way to understand the parity anomaly constraint \constraint.
In this section, we provide a fairly detailed treatment of the regularization of the leading terms in the one-loop effective action for the $U(1)$ gauge field coupled to the fermionic degrees of freedom of a topological insulator.
Our goal is to discuss how the choice of regulator can in general affect the form of this action.
In particular, we show that the effective action has half-integral ${N_f \over 2} + k$ in the limit that the bulk gap $m_0 \rightarrow \infty$.

In general, such a calculation has both IR and UV divergences.
To make the calculation well defined in the IR, we endow the localized fermions with a mass that is small compared to the bulk gap.
The mass acts as an IR regulator and allows us to completely integrate out all matter fields.
However, the mass breaks $P$ and $T$, and generally induces a CS term on the domain wall at level 1/2 (in the limit that the domain wall mass is vanishingly small compared with the bulk gap).
It is important to consider whether or not this level is modified by the regularization procedure.
We show explicitly how the UV divergences of the boundary or defect theory can be removed without a shift of the CS level, and so there is no parity anomaly.
In particular, in the limit of a vanishing $P$ and $T$ violating mass for the boundary modes, no CS term is generated by the regularization procedure -- there is no zero magnetic field Hall conductance.
This is to be contrasted with the situation of a purely 2+1d theory.

Because the section is rather technical, let us provide a brief overview.
We begin by recalling the situation in 2+1d where the removal of UV divergences in a gauge-invariant manner results in the parity anomaly.
After this review, we turn to the calculation of the domain wall effective action.
This proceeds in two steps.
We first calculate the propagator for the modes localized to the surface.
It turns out that while the propagator behaves in the IR as one would expect for a localized Dirac fermion, the UV behavior is softer, decaying faster at large momentum. 
Consequently, its contribution to the effective action remains finite as the UV cutoff is taken to infinity.  (A more careful justification of this claim can be found in Appendix E).  
The bulk is crucial for this effect since the localized modes eventually may mix with the bulk continuum if they are excited to energies comparable to the bulk gap.
Given this behavior, it is then possible to show that a gauge-invariant $P$ and $T$ preserving regularization can be chosen.
While the discussion concerning the regularization of the theory is technical, it has a clear physical interpretation: when $P$ and $T$ are locally broken on the surface, our result makes it clear why a level $1/2$ CS term is obtained for the $U(1)$ gauge field as opposed to an integral level (in the limit of vanishing $\sigma/m_0$).

\subsec{2+1d Regularization Review}

Let us now briefly review how the parity anomaly can be understood in the context of the regularization of QED3 \RedlichDV.
We will have occasion to make reference to this calculation later.
Begin with the action,
\eqn\quantthree{S = \int d^3x\ \bar{\chi} \Big(i \tilde{\gamma}^a (\partial_a - i e A_a + \sigma \Big) \chi,
}
where $\chi$ is a single 2-component Dirac spinor of mass $\sigma$, $\tilde{\gamma}^a = (\sigma^3, - i \sigma^2, i \sigma^1)$ are 2+1d Dirac matrices.
Generally, we expect the regularized effective action for $A_a$ at energies less than $\sigma$ to be a sum of a Maxwell and CS term.

The effective action is found by calculating the fermion determinant obtained by integrating out the fermions.
It is sufficient to consider the leading quadratic terms in $e$ (and consequently in $A$ as well) in the expansion of the determinant.
Working in momentum space,
\eqn\effa{\eqalign{S_{F}(A) = & e^2 \int {d^3 q \over (2\pi)^3} A_a(-q) \Pi^{a b}(q) A_b(q) \cr
= & {e^2 \over 2} \int {d^3 q \over (2\pi)^3} A_a(-q) A_b(q) \int {d^3 p \over (2 \pi)^3} \tr\Big( \tilde{\gamma}^a {i \over (p_c \tilde{\gamma}^c + \sigma)} \tilde{\gamma}^b {i \over (p+q)_d \tilde{\gamma}^d + \sigma}\Big).
}}
The kernel $\Pi^{ab}(q)$ or gauge boson self-energy is UV divergent by power counting.

Thus, we arrive at the delicate (and technical) question of how we should regularize $\Pi^{ab}$.
The general prescription is to choose a regularization that preserves as many of the symmetries present in \quantthree\ as possible.
The usual choice of dimensional regularization does not result in a gauge-invariant effective action; no light fermions are present to compensate for the level $1/2$ CS term present in the resulting action.
The discussion in \S3 demonstrates that the CS level must be integral in a purely 2+1d gapped theory in order that \quantthree\ be invariant under large gauge transformations.
We are interested in maintaining this invariance in our theory so we must choose a different regulator.

The next (archaic) choice is Pauli-Villars.
Here, we first impose an UV cutoff $\Lambda$ on the momentum integral determining $\Pi^{ab}$ to find
\eqn\kernel{\Pi^{ab}(q) = c_1 \Lambda \eta^{a b} + {\sgn(\sigma) \over 8 \pi} \epsilon^{a b c} i q_c + {c_2 \over \sigma} (q^2 \eta^{a b} - q^a q^b) + {\cal O}(1/\Lambda),
}
where $c_1, c_2$ are non-zero, finite constants.
The UV cutoff can be taken to be inversely proportional to an underlying lattice spacing.
The first term is the UV divergence which manifests itself in a gauge non-invariant mass for $A_a$.
Such a term generally appears when a strict cutoff is applied to loop integrals in a gauge theory.
(Dimensional regularization simply sets $c_1 = 0$.)
The second and third terms are a level $1/2$ CS term and the Maxwell term, respectively.

Next, we introduce a single Pauli-Villars field coupled to $A_a$.
The propagator for this field has the same form as that for $\chi$ except for the replacement $\sigma \leftrightarrow M$ with $\sigma \ll M$.
Additionally, the Pauli-Villars field is taken to have bosonic statistics.
Thus, its contribution to $\Pi^{ab}$ is identical in structure to \kernel\ except for an overall sign change of all terms.
Adding these two contributions together gives,
\eqn\kernel{\Pi^{ab}(q) = {\sgn(\sigma) - \sgn(M) \over 8 \pi} \epsilon^{a b c} i q_c + c_2{M - \sigma \over \sigma M} (q^2 \eta^{a b} - q^a q^b) + {\cal O}(1/\Lambda).
}
We can now take the cutoff $\Lambda \rightarrow \infty$.
Notice, however, that decoupling the Pauli-Villars field leaves behind a non-zero contribution to the CS level.
This ``spur" is the manifestation of the parity anomaly.
Parity (and time-reversal) are anomalous in the sense that if $\sigma \rightarrow 0$, the starting action is classically $P$ and $T$ invariant.
Maintaining gauge invariance in the quantum or regularized theory results in the breaking of $P$ and $T$ if we are to use the same regulator for both the massless and massive theories.

Within this regularization scheme, it is not possible to remove by some clever choice of Pauli-Villars fields the gauge non-invariant UV divergence without a non-zero shift of the CS term.
This conclusion is in complete agreement with the topological argument of the previous section and we will, therefore, adopt Pauli-Villars as our regularization prescription.
That is, the requirement of invariance of the low energy action under large gauge transformations is satisfied within the Pauli-Villars regularization scheme, but not within dimensional regularization.

The lesson is that there is a physical difference between dimensional and Pauli-Villars regularization in 2+1d.
If invariance of the theory under large gauge transformations is to be maintained, then Pauli-Villars regularization must be used and parity is broken.
From the discussion of the previous section, we anticipate that there is no physically observable difference between Pauli-Villars and dimensional regularization when the theory lives on the boundary of a contractible higher-dimensional space because there do not exist large gauge transformations that extend continuously into the bulk in any such theory.
In the next section, we show explicitly how this equivalence comes about.

\subsec{Propagator}

We now turn to the calculation of the regularized effective action for the localized modes described at low energies by \quantthree.
Lest we run into a contradiction with the conclusion of \S3, the action \quantthree\ must only be a low energy approximation to the physics.
We will find that the 3+1d bulk, which is implicit in any low energy domain wall action, will play an essential role.

In order to derive the action, we first determine the full propagator for the domain wall modes by generalizing the nice work of Chandrasekharan \ChandrasekharanAG.
Perhaps surprisingly, the form of the propagator is partially determined by the massive bulk modes. 
Our derivation is contrasted with an anomaly inflow argument in Appendix B.
We have also repeated in Appendix D the calculations in this section for the analogous system in the simpler 1+1d context.

As we reviewed in \S2, we describe the 3+1d topological insulator by the action,
\eqn\startaction{S = \int d^4x\ \bar{\psi}\Big(i \gamma^{\mu}(\partial_\mu - i e A_\mu) + m(x) + i \gamma^5 \sigma \Big)\psi,
}
where $\psi$ is a 4-component spinor that describes the massive bulk fermion with spatially dependent mass $m(x)$ coupled to the $U(1)$ gauge field $A_\mu$ which may represent the electromagnetic field.
The spatially-varying mass allows us to study the interface between a topologically trivial and topologically non-trivial insulator, while the constant T-breaking mass $\sigma$ regularizes the infrared divergences on the domain wall.  
(We have not included kinetic terms for the gauge field, since their specific form does not affect our results.) 
Note that $\bar{\psi} = \psi^{\dagger} \gamma^0$.

We take $m(x_3)$ to depend only on a single coordinate $x_3$ and to have the profile, $m(x_3) = m_0 \tanh(x_3/\ell)$ with $m_0 > 0$.
If $\sigma \neq 0$, we may safely integrate out $\psi$ without an IR divergence.
We assume that $0 < \sigma \ll m_0$.
(In the opposite limit, there is no domain wall and the entire system is in a single massive $P$ and $T$ breaking phase).

We are interested in the one-loop effective action for $A_\mu$ obtained after integrating out $\psi$.
This action takes the generic form,
\eqn\actionefffour{S = - {e^2 \over 2} \int d^4 x d^4 x' A_\mu(x) A_\nu(x') \tr\Big(\gamma^\mu D(x,x') \gamma^\nu D(x',x)\Big),
}
where $D(x,x')$ is the fermion propagator and the overall minus sign comes from the fermion loop.
We are particularly interested both in divergent terms which are implicit in \actionefffour\ that require regularization and in possible CS terms localized to the 2+1d domain wall. 
These CS terms may arise either directly from modes localized to the domain wall or through an integration by parts of a bulk $\theta$-term.

Thus, our first task is to calculate $D(x,x')$. 
We choose the following representation for the Dirac matrices,
\eqn\diracrep{\gamma^0 = - i \pmatrix{0 & 1 \cr -1 & 0},\ \gamma^i = - i \pmatrix{\sigma^i & 0 \cr 0 & - \sigma^i}, \gamma^5 = i \gamma^0 \gamma^1 \gamma^2 \gamma^3 =  \pmatrix{0 & 1 \cr 1 & 0}.
}
The eigenspinors of the Dirac equation at $e = 0$,
\eqn\eigeneqn{\gamma^0 \Big(i \gamma^\mu \partial_\mu + m(x_3) + i \gamma^5 \sigma \Big) \psi^\lambda = \lambda \psi^\lambda.
}
formally define the propagator,
\eqn\propagator{D(x,x') = i \sum_\lambda {\psi^{\lambda}(x) \bar{\psi}^{\lambda}(x') \over \lambda},
}
where the eigenfunctions $\psi^\lambda$ are normalized with respect to the usual inner product,
\eqn\innerprod{\int d^4 x (\psi^\lambda)^{\dagger}(x) \psi^{\lambda'}(x) = \delta({\bf \lambda} - {\bf \lambda'}).
}

Translation invariance in the temporal and spatial directions parallel to the domain wall allows us to express the propagator in a mixed Fourier space representation,
\eqn\mixedprop{D(k, x_3, x_3') = i \sum_{\lambda} {\psi_k^{\lambda}(x_3) \bar{\psi}_k^\lambda(x_3') \over \lambda},
}
where $k$ collectively refers to momentum parallel to the wall and $\psi_k^\lambda$ are the Fourier coefficients in the expansion of $\psi^\lambda$.
Expanding in Fourier modes in this mixed basis, the Dirac equation becomes,
\eqn\diracfourier{\gamma^0 \Big( \gamma^a k_a + i \gamma^3 \partial_3 + m(x_3) + i \sigma \gamma^5 \Big) \psi_k^\lambda = \lambda \psi_k^\lambda,}
where $a = 0,1,2$.
Two types of eigenstates are expected: modes localized near the domain wall and modes allowed to propagate away from the domain wall.
We refer to the former as bound modes and the latter set as scattering modes.

Consider first the scattering modes.
The eigenvalues for the scattering states may be obtained by considering \diracfourier\ in the limit $x_3 \rightarrow \pm \infty$.
In this limit, translation invariance along the $x_3$-direction is effectively restored and so we introduce an asymptotic momentum $k_3$ parametrizing the eigenvalues of \diracfourier.
There are two pairs of eigenspinors $\psi^{\lambda(k_3) \pm, (i)}_{k}$ with eigenvalues, $\lambda(k_3)_{\pm} \equiv \lambda_{\pm} = k_0 \pm \sqrt{k_1^2 + k_2^2 + k_3^2 + m_0^2 + \sigma^2} = k_0 \pm \omega_k$.

The coupled first order equation \diracfourier\ may be rewritten as the second order equation,
\eqn\fourdiracalt{\Big(\partial_3^2 + k_3^2 + m_0^2(1 + {i \gamma^3 \over m_0 \ell}) \sech^2({x_3 \over \ell}) \Big) \psi_k^{\lambda_{\pm}} = 0.}
Above, we have substituted the mass profile $m(x_3) = m_0 \tanh(x_3/\ell)$.
This equation says that each component of $\psi_k^{\lambda_{\pm}}$ satisfies a modified P\"oschl-Teller equation, with parameter $m_0/\ell$.
The associated eigenspinors and eigenvalues to this equation are known for general $m_0/\ell$, however, they can be given simple closed form expressions when $\ell = 1/m_0$.
For this reason, we specialize to this point in parameter space for the remainder of the paper.

There is a well known connection between the P\"oschl-Teller equation and supersymmetric quantum mechanics (see e.g., \stonegoldbart).
The upshot is that the separation of eigenspinors into bound and scattering states can be made precise.
Eigenspinor components of scattering states are paired; while those of the bound states are zero modes of the associated supersymmetric charge operator.
The number of such zero modes depends upon the ratio $\ell/m_0$ which we set to unity, thereby implying a single set of bound state modes.
See Appendix D for further details about this connection in the technically simpler case of 1+1d.

An orthonormal basis for the scattering state solutions to \fourdiracalt\ is provided by
\eqn\scatfouralt{\eqalign{\psi^{\lambda_{\pm},(1)}_k(x_3) = &  {1 \over {\cal N}_{\pm}^{(1)}} \pmatrix{k_3 - i m(x_3) \cr k_1 + i k_2 \cr \pm \omega_k - \sigma \cr 0} e^{- i k_3 x_3}, \cr
\psi^{\lambda_{\pm},(2)}_k(x_3) = & {1 \over {\cal N}_{\pm}^{(2)}} \pmatrix{(k_1 - i k_2)(k_3 - i m(x_3)) \cr - k_3^2 - m_0^2 \cr 0 \cr (\pm \omega_k - \sigma)(k_3 - i m(x_3))} e^{- i k_3 x_3},
}}
where
\eqn\normalizations{{\cal N}_{\pm}^{(1)} = \sqrt{2 (2 \pi)^4 \omega_k (\omega_k \mp \sigma)}, \quad {\cal N}_{\pm}^{(2)} = {\cal N}_{\pm}^{(1)} \sqrt{k_3^2 + m_0^2}.
}

It remains to find the bound states.
A bound state takes the form $\psi^{\lambda_0}_k = \pmatrix{a \chi_1 & 0 & 0 & b \chi_2}^{\tr} \sech(m_0 x_3)$.
These states do not carry ``momentum" $k_3$; they are localized to the domain wall because of the prefactor $\sech(m_0 x_3)$.
A normalized basis is provided by 
\eqn\boundfour{\psi^{\lambda_0,\pm}_k(x_3) = \sqrt{{m_0 \over 4 (2 \pi)^3 \epsilon_k (\epsilon_k \mp \sigma)}} \pmatrix{k_1 - i k_2 \cr 0 \cr 0 \cr \pm \epsilon_{k} - \sigma} \sech(m_0 x_3),
}
with eigenvalues $k_0 \pm \epsilon_k = k_0 \pm \sqrt{k_1^2 + k_2^2 + \sigma^2}$.

Having found the eigenspinors and eigenvalues, we can now construct the propagator using \propagator.
It is a sum of two terms arising from the bound and scattering modes,
\eqn\fullprop{D(k, x_3,x_3') = D^{{\rm bound}}(k, x_3,x_3') + D^{{\rm scat}}(k, x_3,x_3').
}
We compactly write the contribution to the propagator from the scattering states as
\eqn\scatpropthree{D^{{\rm scat}}(k, x_3,x_3') = i \int {dk_3 \over 2\pi} {\gamma^\mu k_\mu + M(x_3,x_3') + i \sigma \gamma^5 \over k_a^2 - k_3^3 - m_0^2 - \sigma^2} e^{i k_3 (x_3' - x_3)},
}
for $\mu = 0, ..., 3$ and where $k_a^2 = k_0^2 - k_1^2 - k_2^2$.
The unconventional mass matrix,
\eqn\massexpand{M(x_3, x_3') = - {m(x_3) \over 2} (1 + i \gamma^3) - {m(x_3') \over 2}(1 - i \gamma^3) + {1 \over 2} (1 + i \gamma^3) {(\gamma^a k_a + i \sigma \gamma^5) \over (k_3^2 + m_0^2)} \mu(x_3, x_3').
}
and
\eqn\mumatrix{\mu(x_3, x'_3)_{k_3} = m(x_3) m(x'_3) + i k_3 \Big(m(x'_3) - m(x_3)\Big) - m_0^2.}
Notice that $M(x_3, x_3')$ approaches the standard form $- m_0 $ if the mass $m(x_3)$ is taken to be a constant.
The bound state contributes the following term to the propagator,
\eqn\boundpropthree{D^{{\rm bound}}(x,x') = {i \over 2} {m_0 \over 2} (1 + i \gamma^3) \sech(m_0 x_3) \sech(m_0 x'_3) {\gamma^a k_a + i \gamma^5 \sigma \over k_a^2 - \sigma^2} }
for $a = 0, 1, 2$.

\subsec{Corrections to the Effective Action}

We now use the above propagator to study corrections to the gauge field effective action.
Instead of plugging the full propagator into \actionefffour, it  is useful to first study the structure of the propagator itself more closely.

Before doing so, two technical comments are in order.
First, we have been working in Lorentzian signature in the previous sections, however, we have found it most convenient to perform the necessary (intermediate step) integrals below by first Wick rotating to Euclidean signature by substituting $k_0 \rightarrow i k_0$ so that $k_a^2 \rightarrow - |k_a^2|$.
Our resulting expressions, however, are written with the original Lorentzian metric. 
Second, we use a renormalization procedure in which we integrate over all $k_3$, but impose an UV cutoff $\Lambda$ on the remaining three momentum integrals $k_a$.  This unconventional choice should not affect the low energy properties of the regularized theory.

$D^{{\rm bound}}$ is the Fourier transform of the usual 2+1d Dirac propagator.
Indeed, despite the slightly different matrix structure, substituting $D^{{\rm bound}}$ into \actionefffour\ gives precisely the same contribution to the effective action as a truly 2+1d Dirac fermion reviewed in \effa.
The hyperbolic prefactors merely localize the contribution to the domain wall.
The localization becomes exact in the $m_0 \rightarrow \infty$ limit where we use the relation
\eqn\approxdelta{\lim_{m_0 \rightarrow \infty} {m_0 \over 2} \sech^2(m_0 x_3) = \delta(x_3).
}
Therefore, we must understand how the UV divergences are regularized.
In particular, we wish to determine whether or not there is a shift to the CS level if we regularize using a Pauli-Villars scheme.

A clue comes from a closer inspection of the full propagator.
Not only do the bound modes result in propagation that is localized to the domain wall; surprisingly, the scattering modes contribute a term that is localized as well!

The unconventional terms in the mass matrix $M(x_3, x_3')$ proportional to $\mu(x_3, x_3')$ are responsible for this localization.
Let us momentarily focus upon these terms in the scattering mode contribution \scatpropthree\ to the propagator proportional to $\mu(x_3, x_3')$.
We denote these terms by $D_{\mu(x_3)}^{{\rm scat}}$.
After a contour integration over $k_3$,  they take the form 
\eqn\scatlocal{\eqalign{
D_{\mu(x_3)}^{{\rm scat}}(k, x_3, x_3') = - D^{{\rm bound}}(k, x_3, x_3') - {i \mu(x_3, x_3')  \over 4} (1 + i \gamma^3)  {(\gamma^a k_a + i \sigma \gamma^5) \over (k_a^2 - \sigma^2)} {e^{ i k_3  (x_3' - x_3)} \over  \sqrt{- k_a^2 + m_0^2 + \sigma^2}},
}
} 
where $k_3$ above is evaluated at $k_3 = i \sgn(x_3' - x_3) \sqrt{- k_a^2 + m_0^2 + \sigma^2}$. 

Remarkably, the scattering modes contribute a term that is equal and opposite to the contribution to the propagator arising from the bound modes.
These two terms cancel one another in the full propagator.
The remaining localized term in the propagator is the second term appearing in \scatlocal.
Thus, we define the localized propagator,
\eqn\loc{D^{{\rm loc}}(k, x_3, x_3') = - {i \mu(x_3, x_3')  \over 4} (1 + i \gamma^3)  {\gamma^a k_a + i \sigma \gamma^5 \over (k_a^2 - \sigma^2)\sqrt{-k_a^2 + m_0^2 + \sigma^2}} e^{i k_3 (x_3' - x_3)},
}
where, as above, $k_3 = i \sgn(x_3' - x_3) \sqrt{- k_a^2 + m_0^2 + \sigma^2}$. 
This propagator describes the excitations that are restricted to living along the domain wall at energies low compared to the bulk gap. 
This is the main technical result of our paper and it is the form of this localized propagator that is the key allowing us to understand the divergence structure of the 2+1d boundary theory.

Notice that the UV behavior of $D^{{\rm loc}}$ is softer than $D^{{\rm bound}}$: it decays faster as $k\rightarrow \infty$. 
This behavior implies that the purely local mode contribution to the effective action \actionefffour\ is finite as the UV cutoff is taken to infinity.
(By purely local mode contribution to the effective action, we mean \actionefffour\ with $D^{{\rm loc}}$ substituted in place for the general propagator $D$.)
We shall begin by calculating these finite terms.
Afterwards, we comment upon the corrections to the action arising from interactions between modes localized to the domain wall and those modes that are free to propagate throughout the bulk.
A detailed analysis of these latter two terms is relegated to Appendix E.

To isolate the finite contributions to the action from the localized modes, we find it convenient to use the following approximate expression for the local propagator,
\eqn\localapprox{D^{{\rm loc}}_{{\rm approx}}(k, x_3, x_3') = - {i \over 2} (1 + i \gamma^3) {m_0^2 \over 2} \sech(m_0 x_3) \sech(m_0 x_3') {\gamma^a k_a + i \sigma \gamma^5 \over (k_a^2 - \sigma^2) \sqrt{- k_a^2 + m_0^2 + \sigma^2}}.
}
This expression becomes a better approximation to the exact result \scatlocal\ as the limit $m_0 \rightarrow \infty$ is approached.
We expect corrections to this approximation to be suppressed in the large bulk mass limit.  
(In Appendix E, we evaluate the contribution of the localized mode without making the above approximation and find that \localapprox \ gives the correct qualitative structure for the divergences and the quantitatively correct value for any finite terms).  

Using the approximate local propagator \localapprox, we may now compute its contribution to the gauge field effective action.
In particular, the kernel or one-loop self-energy is
\eqn\locsecond{\Pi_{ab}(q, x_3 x_3') = \Big(- {|m_0| \over 6}  \eta^{c d} + {\sgn(\sigma) \over 8 \pi} i q_b \epsilon^{c b d} + {c_2 \over |\sigma|} (q^2 \eta^{a b} - q^a q^b) + {{\cal O}}({1 \over \sigma^2}, {1 \over m_0})\Big) \delta(x_3) \delta(x_3'),
}
where $c_2$ is a finite non-zero constant and the delta functions arise from the hyperbolic prefactors using \approxdelta.
We stress that the above result is finite due to the faster decay of \localapprox\ at large momentum, so that a 2+1-dimensional UV regulator is not required.
However, we have exchanged a gauge non-invariant term proportional to the cutoff for precisely the same term, now proportional to the bulk gap $m_0$.

Therefore, we must understand how to properly decouple the bulk by taking $m_0 \rightarrow \infty$.
It is essential that the divergence is proportional to the bulk gap rather than an UV cutoff.
This difference allows us to regularize the theory by Pauli-Villars fields without shifting the CS level.
(In Appendix E, we show in detail that the linear divergence appearing as $m_0 \rightarrow \infty$ is the only possible divergence that we may associate with presence of the localized states.
${\cal O}(1)$ corrections to the coefficient of this term from the value displayed in \locsecond\ may arise from interactions with the bulk modes, however, the precise value of the coefficient does not affect the arguments below.
Additionally, there are the usual (bulk) divergences associated to QED4.)

We merely need to introduce two Pauli-Villars fields whose propagators have roughly the same form as the localized fermion \localapprox.
The differences lie in the choice of statistics $C_i$ and masses for the Pauli-Villars fields.
We endow the first Pauli-Villars field with bosonic statistics, $C_1 = 1$, and replace $m_0 \leftrightarrow M$ and $\sigma \leftrightarrow \sigma'$.
The second field is taken to have fermionic statistics, $C_2 = -1$, and we replace $m_0 \leftrightarrow M - m_0$ and $\sigma \leftrightarrow \sigma'$.
We assume the hierarchies $m_0 \ll M$ and $\sigma \ll \sigma' < m_0$.
Adding the contributions from the physical boundary fermion and the two Pauli-Villars fields, \locsecond\ becomes
\eqn\locsecondreg{\eqalign{\Pi_{ab}(q, x_3, x_3') = {1 \over 6}(- |m_0| + C_1 |M| + C_2 |M-m_0| 
+ {\sgn(\sigma) + C_1 \sgn(\sigma') + C_2 \sgn(\sigma') \over 8 \pi} i q_b \epsilon^{c b d},
}}
where we have suppressed writing the contributions to the Maxwell term and it is to be understood that the above correction is localized at $x_3 = x_3' = 0$.
Clearly our choice of statistics and masses removes the term in \locsecond\ proportional to $m_0$, but retains the half-integral level of the CS term induced by the localized fermion.
As promised, we have regularized the boundary theory with Pauli-Villars fields in a $P$ and $T$ invariant manner, i.e., we have regularized without shifting the CS level and so no CS term is generated by the regularization procedure when $\sigma \rightarrow 0$.
We were successful in doing so because of the softer UV properties of the localized fermion which resulted in a divergence proportional to the bulk gap rather than the UV cutoff.
This enabled the theory to be regularized with an even instead of an odd number of Pauli-Villars fields.

The form taken by the Pauli-Villars fields' propagators betrays their 3+1d origin. 
Indeed 3+1d QED requires at least three Pauli-Villars fields for its regularization \ItzyksonRH.
We have merely chosen two of the Pauli-Villars fields in the 3+1d bulk to have soliton masses similar in form to the physical fermion so that they can regularize the lower-dimensional theory.
The remaining 3+1d Pauli-Villars fields can be given masses that are positive everywhere. 

The {\it finite} gauge non-invariant photon mass term (which we removed in \locsecondreg) may be surprising. 
However, its appearance is similar to what occurs in the Pauli-Villars regularization of QED4 where a gauge non-invariant mass term is also found \ItzyksonRH.
In that case, the mass squared is a sum of two terms: one proportional to the square of the cutoff and a finite term proportional to the square of the bulk fermion mass.  
In both cases, the appearance of gauge non-invariant terms proportional to a power of the fermion mass is a result of the fact that imposing a momentum cutoff $\Lambda$ (in our prescription for the 3-momenta $k_a^2$, or for all 4-momenta $k_\mu^2$ in the standard QED4 case) breaks gauge invariance explicitly.  
The correct choice of Pauli-Villars regulator fields is determined by the criterion that they must fully restore the gauge symmetry broken by this choice of cutoff.  

Now that we have explained how the excitations described by the local propagator are regularized, we should ask: Can the massive bulk modes contribute non-trivial terms to the gauge field effective action?
By massive bulk modes, we mean the terms in $D^{{\rm scat}}$ that are not localized to the domain wall by any hyperbolic prefactors.  Specifically, the full fermion propagator is given by 
$D =  D^{{\rm loc}} + D^{{\rm free}}$, with  $D^{{\rm loc}}$ given by \loc, and
\eqn\Dfrees{
D^{{\rm free}} =     {i e^{ i k_3  (x_3'-x_3) } \over 2 \sqrt{ - k_a^2 + m_0^2+ \sigma^2}  }   \left(   \gamma^a k_a -  \gamma^3 k_3  - {m(x_3)  \over 2} (1 + i \gamma^3) - {m(x_3') \over 2}(1 - i \gamma^3)  + i \sigma \gamma^5  \right )  \ \  \ .
}
where as above, $k_3 = i \ {\rm sign} (x_3'-x_3) \sqrt{ - k_a^2 + m_0^2+ \sigma^2}$.
The corrections that we have thus far ignored arise from either single insertions of $D^{{\rm loc}}$ and $D^{{\rm free}}$ or two insertions of $D^{{\rm free}}$ into \actionefffour.
These correct both the boundary and bulk Lagrangians.
For example, the bulk Maxwell term is radiatively modified by the massive bulk modes described by $D^{{\rm free}}$, while the cross-term of $D^{{\rm loc}}$ with $D^{{\rm free}}$ modifies both the $2+1$d Maxwell term {\it and} the effective Chern-Simons term arising at the domain wall.

As with the local contribution discussed above, there are two aspects to this question.
First, we must determine if there can be a direct correction to either a bulk $\theta$-term or (what is the same) a boundary CS term.  Second, we must determine what effect (if any) these contributions have on the choice of regulator fields that we must include in the theory.  

It is straightforward to show that the only correction to the boundary CS term arises from the ``crossterm" between $D^{{\rm loc}}$ and $D^{{\rm free}}$.  
No bulk $\theta$-term  is generated.
Adding this crossterm correction to our result \locsecondreg, we find the total CS level,
\eqn\NotQuantised{
 k = {\eta \sigma \over 2 |\sigma|} -{1 \over \pi} \tan^{-1} \Big({\sigma \over \eta m_0}\Big),
}
where $\eta = \pm 1$ is defined by the orientation of the domain wall, $m(x_3) = \eta m_0 \tanh(m_0 x_3)$.
(Our calculation above specialized to the case $\eta = 1$. 
Similar manipulations show that \NotQuantised\ obtains for $\eta = -1$.)
Precisely the same non-quantized correction to the CS level also occurs in the analogous 1+1d situation -- see (D.23).
(In 1+1d, the CS level directly determines the induced charge on the soliton, however, in 3+1d the soliton only carries a charge when a background magnetic field is applied perpendicular to the surface.)
Extrapolating the 1+1d intuition (axion electrodynamics is suggestive as well \WilczekMV), we expect that for general (not necessarily constant) $\sigma$,
\eqn\kphase{
k = {\Delta \phi \over 2\pi},
} 
where $\Delta \phi$ is the change in phase $\phi$ of the complexified bulk fermion mass during any interpolation. 
The fact that \kphase\ need not be rational is not in conflict with gauge invariance; for a single domain wall, there is no quantization condition on the CS level (as shown in \S3).  
For a system with two domain walls, the phase of the fermion mass must wind by an integer multiple of $2 \pi$ as it crosses {\it both} boundaries, provided that the fermion mass asymptotes to the same value everywhere outside the system.

Are there divergences appearing as the cutoff $\Lambda \rightarrow \infty$ that must be regularized?
We study these possible divergences in Appendix E and show that they do not occur. 
There are two types of possible divergences.
The first kind arise from the usual massive 3+1d propagator and are familiar from QED4. 
Such divergences are always present and can be regularized (with at least three Pauli-Villars fields \ItzyksonRH) without changing the effective CS level.
The second type of divergence is special to the introduction of a domain wall mass profile and is localized along the wall.
Such divergences are at worst logarithmic in $\Lambda$; 
however, a precise cancellation occurs between the bulk and localized modes, described above, such that only finite terms remain as $\Lambda \rightarrow \infty$.
We stress that this cancellation is important: a divergent result would have required adding an {\it odd} number of regulator fields with domain-wall mass profiles, thereby rendering the CS level $k$ an integer (up to the corrections of order $\sigma/m_0$ noted above), forcing the constraint \constraint\ to be obeyed in the limit $\sigma/m_0 \rightarrow 0$.

\newsec{Conclusion}

In this work, we have examined the relationship between the parity anomaly and the gapless Dirac fermion arising at the surface of a 3+1d topological insulator. 
Naively, a coupling of the localized surface mode to a fluctuating gauge field would result in the parity anomaly on each boundary surface, i.e., the requirement that ${N_f \over 2} + k$ be integral on each boundary surface where $N_f$ is the number of fermions localized at the boundary, and $k$ is the sum of any bare or induced CS level.
We have shown that this does not occur and that ${N_f \over 2} + k$ is half-integral in the limit that the bulk gap $m_0 \rightarrow \infty$.
Hence, an odd number of gapless Dirac fermions on any such surface can be coupled to fluctuating (bulk) gauge fields and still maintain parity and time-reversal symmetries.

We have come to this conclusion using two complementary perspectives: topological quantization conditions and regularization.
It is invariance of a 2+1d effective theory under large gauge transformations that results in the constraint that ${N_f \over 2} + k$ be integral.
When the system of interest lives on the boundary of a higher-dimensional space, such large gauge transformations either do not exist (as they cannot be extended continuously into the bulk) or they are innocuous -- their effect is cancelled between all components of the boundary.  Thus the topological quantization conditions that gauge invariance imposes on 2+1d theories do not apply in 3+1d.
Similarly, Pauli-Villars regularization of QED3 preserves the invariance of the theory under large gauge transformation at the cost of breaking parity and time-reversal invariance. 
We have shown explicitly how for a single species of Dirac fermion on the 2+1d boundary of a 3+1d bulk, the presence of the bulk softens the UV properties of the boundary fermions, such that the theory admits 
a parity and time-reversal invariant Pauli-Villars regularization.
 
It is instructive to contrast our result with the quantum Hall effect.
Here, invariance under local gauge transformations (charge conservation) requires that both a bulk CS term and boundary chiral excitations be present in a low energy description of the system.
In contrast, only a global anomaly can be present for the surface modes of a 3+1d topological insulator.
However, the fact that the gauge field is free to explore the bulk essentially eliminates the large gauge transformations responsible for the possible global anomaly; there is no mutual cancellation of anomalous transformations as occurs in the quantum Hall effect.  Thus unlike the chiral edge modes of a quantum Hall system, the gapless boundary modes of a 3+1d topological insulator are not required to preserve gauge invariance.

The fact that there are no topological quantization conditions for these boundary theories has important physical consequences.  
A purely 2+1d system exhibiting a fractional Hall conductivity of $\sigma_{xy} = {1 \over p} {e^2 \over h}$ with integer $p$ can be gauge invariant only if the system exhibits a ground state degeneracy equal to $p$ when placed on a spatial torus.
Otherwise, the theory fails to be invariant under large gauge transformations generated by flux insertions through the two non-trivial cycles of the torus.
If we fill in the center of the torus, however, the non-trivial large gauge transformations that cause this problem no longer exist (in the sense that they do not continuously extend into the bulk), and
the requirement of a degenerate ground state disappears.
Hence, as we would expect for a non-interacting system, the ground state of a topological band insulator is unique in spite of the fact that $p=2$ on its surface (in the limit of vanishing $\sigma/m_0$ with finite $\sigma$).  (For a fractional topological insulator in 3+1d, there is a ground state degeneracy, but this arises from its bulk topological order\SwingleEtAl, and not from the surface).   If instead we thicken the torus, gauge invariance requires a ground state degeneracy that is half of what one might naively expect based on the Hall conductivity, since the large gauge transformations must behave identically on both surfaces.  For the topological band insulator this again implies a unique ground state.

In fact, the perturbative calculation reveals  that when both bulk and boundary fermionic mode contributions are considered, the coefficient of the Chern-Simons term (physically, the Hall conductivity at the surface) is {\it not} quantized to be a rational fraction when ${\sigma \over m_0}$ is non-vanishing.  
Rather, it is of the form
\eqn\NotQuantised{
 k = {\eta \sigma \over 2 |\sigma|} -{1 \over \pi} \tan^{-1} \Big({\sigma \over \eta m_0}\Big),
}
where $\eta = \pm 1$ determines the orientation of the domain wall via the soliton mass profile, $m(x) = \eta m_0 \tanh(m_0 x_3)$.
Here, $m_0$ is the bulk band gap, and $\sigma$ is the time-reversal and parity breaking mass of the boundary fermions.
For a general soliton mass profile, we expect the CS level to equal $1/2\pi$ multiplied by the total change in phase of the complexified fermion mass.
(In Appendix D, we show that the CS level for an analogous 1+1d system is also given by \NotQuantised.)
We may think of the half-integral contribution coming from the fermionic modes localized to the boundary, while the non-quantized contribution arising from the interaction between the bulk fermions and the modes localized to the boundary.  

In a real system where the ratio $\sigma/m_0$ is not asymptotically vanishing, the above deviation from half-integrality may be observable. 
A magnetic-susceptibility measurement would in-principle measure the difference in the Hall conductivities.
Combined with a Kerr or Faraday rotation measurement (which measures the sum), one could then extract the Hall conductivities of each boundary surface \tsemacone. 
However, we caution that strict use of the formula \NotQuantised\ requires a non-vanishing time-reversal breaking perturbation to be present at asymptotically large distances from any topological insulator boundary.

We stress that in order for our conclusions to be valid, the gauge field must be free to explore the bulk of the system.
It is possible to imagine a situation in which strong correlations among the modes localized to the boundary of the system lead to fractionalization. 
If fractionalization only occurs at the boundary, it can be described by an ``emergent" gauge field that only has support on the lower-dimensional boundary and so any resulting constraints imposed by gauge invariance truly are of a lower-dimensional origin.
An alternative scenario for localizing a truly 2+1d gauge field uses a bulk Higgs field charged under some gauge group whose symmetry-breaking profile only allows a subgroup (possibly, an empty one) of light ``photons" in the bulk, but  a domain wall defect where the full gauge symmetry is restored.
\foot{We thank S. Kachru for discussions on this latter possibility.}  
We hope to discuss these scenarios further in future work.

\bigskip
\medskip
\centerline{\bf{Acknowledgements}}

It is a pleasure to thank Allan Adams, Eduardo Fradkin, Michael Freedman, Shamit Kachru, John McGreevy, Joel Moore, and Chetan Nayak for useful discussions and comments on a draft of this paper. 
M.M. acknowledges the generous support and hospitality of the Center for Theoretical Physics at MIT during the beginning stages of this work and the Aspen Center for Physics and the NSF Grant $\#1066293$ during its conclusion.  F. J. B. is thankful to the hospitality of KITP (NSF PHY11-25915) during part of this collaboration.

\appendix{A}{Domain Wall Fermions}

In this appendix, we review the domain wall fermions \refs{\JackiwFN, \KaplanBT} that arise in the model,
\eqn\fouraction{S = \int d^4 x \Big( \bar{\psi} i \gamma^\mu (\partial_\mu - i e A_\mu) \psi + m(x) \bar{\psi} \psi \Big),
}
where  the mass $m(x)$ is real and satisfies,
\eqn\asymptotic{\lim_{x_3 \rightarrow \pm \infty} m(x_3) = \pm m_0, \ \ m_0 >0}
with $m(x_3)$ passing through zero exactly once, at $x_3=0$.
We can think of these gapless domain wall fermions as zero modes bound to a defect in the order parameter where symmetry is restored. 
In this example, the order parameter is the real field $m(x_3)$ and the symmetry that is (classically) restored at the defect or location where $m(x_3)$ vanishes is the chiral symmetry, $\psi \rightarrow \exp(i \alpha \gamma_5) \psi$.
(This is only a true symmetry of the quantum theory at zero gauge coupling due to the chiral anomaly \refs{\adler,\belljackiw, \fujikawa}.)

To see how these zero modes arise, it is useful to write the Dirac equation as,
\eqn\dequation{\Big(i D_{||} + i D_\perp + m(x_3)\Big)\psi = E \gamma^0 \psi,
}
where 
\eqn\dops{\eqalign{ D_{||} = & \gamma^a (\partial_a - i e A_a),\ a=0,1,2 \cr
D_\perp = & \gamma^3 (\partial_3 - i e A_3).
}}
A zero mode satisfies the equation,
\eqn\zequation{\Big(i D_{\perp} + m(x_3)\Big) \psi = E \gamma^0 \psi,
}
with $E = 0$.
We are interested in finding a solution about the free theory so we may set $e=0$ in the equations of motion.

It is convenient at this point to choose a particular representation for the Dirac matrices. 
The results, however, are independent of any particular choice.
We take
\eqn\diracs{\gamma^0 = \pmatrix{0 & {\bf 1}_{2 \times 2} \cr {\bf 1}_{2 \times 2} & 0},\ \gamma^i = \pmatrix{0 & \sigma^{i} \cr - \sigma^i& 0},
}
where $\sigma^i$ for $i = 1,2,3$, are the usual Pauli-sigma matrices. 
Note that the above choice is a rotation from that used in \S4.

Since $\tr(\gamma^3) = 0$ and $(\gamma^3)^2 = - \bf{1}_{4 \times 4}$, $\gamma^3$ possesses two pairs of eigenvalues equal to $\pm i$ with eigenspinors defined by the equation
\eqn\eigenspin{\gamma^3 \psi_{\pm} = \pm i \psi_\pm.}
A solution to \eigenspin\ contains half the number of degrees of freedom of a 3+1d Dirac fermion.
The eigenspinors take the form,
\eqn\zeromodeans{\psi_{\pm}(x) = \phi_{\pm} \pmatrix{ \chi \cr \pm i \sigma^3 \chi},
}
where $\chi(x)$ is an arbitrary two-component spinor that depends only on the three coordinates parallel to the domain wall and 
\eqn\psol{\phi_{\pm}(x_3) = C_{\pm}  \phi_{\pm}^{(0)} = C_{\pm} \exp\Big( \pm \int_{x_3^{(0)}}^{x_3} m(x_3)\Big).}
$C_{\pm}$ is a normalization constant and $x_3^{(0)}$ can be chosen to coincide with the location of the domain wall.
In the following analysis, it is convenient to normalize $\psi_{\pm}$ by choosing $C_{\pm}^{-2} = 2 \int (\phi_{\pm}^{(0)})^2 dx_3$.
While both $\psi_\pm$ solve the zero mode equation, only one is normalizable; only one has a finite, non-zero $C_{\pm}$.
The asymptotics \asymptotic\ chosen above for $m(x_3)$ singles out $\psi_-$ as the normalizable zero mode. 
Had the opposite asymptotics been chosen, $\psi_+$ would be the normalizable zero mode. 

Now consider the action that describes these fermionic localized modes at energies much less than the bulk band gap, $m_0$.
It is found by substituting $\psi_\pm$ into the 3+1d action \fouraction.
Given the profile \asymptotic for $m(x_3)$, we plug $\psi_-$ into the action to find
\eqn\leftaction{\eqalign{S(\psi_-) = & \int d^4x \Big( \bar{\psi}_- i \gamma^\mu (\partial_\mu - i e A_\mu) \psi_- + m(x_3) \bar{\psi}_- \psi_- \Big) \cr
= & C_-^2 \int dz\ (\phi_-^{(0)})^2 \int d^3x\Big( \pmatrix{\chi^\dagger & - i (\sigma^3 \chi)^\dagger} \gamma^0 \gamma^a \Big(\partial_a - i e A_a\Big) \pmatrix{\chi & i \sigma^3 \chi}^T \Big) \cr
= & \int d^3 x \Big(\bar{\chi}\ i \tilde{\gamma}^a (\partial_a - i e A_a) \chi\Big).
}}
The above action describes a massless 2-component Dirac fermion coupled to a $U(1)$ gauge field, namely QED3.
Again, we have suppressed the tree-level kinetic term for $A_\mu$ restricted to the domain wall.
It is sufficient to say that the tree-level gauge boson propagator restricted to the 2+1d surface is softer in the IR, decaying as $1/|p|$ as opposed to $1/p^2$ at small momentum, because of an integration over the direction normal to the domain wall.
The resulting 2+1d Dirac matrices are
\eqn\threedirac{\tilde{\gamma}^a = \pmatrix{\sigma^3, - i \sigma^2, i \sigma^1},\ a = 1,2,3.
}
Note that the coefficient of the minimal coupling term in the action between the fermion number current in the $x_3$-direction and $A_3$ vanishes identically since $\bar{\psi}_\pm \gamma^3 \psi_\pm = 0$.
This ensures that no zero modes leak off the defect via a coupling to $A_3$. 

\appendix{B}{Anomaly Inflow}

In this appendix, we first review anomaly inflow for the case of a string defect in 3+1d.
We then contrast this analysis to that of a domain wall defect in 3+1d which is relevant to this paper.

The original model studied by Witten \WittenEB\ and Callan and Harvey \CallanSA\ is 
\eqn\startactionfour{S = \int d^4 x\ \bar{\psi}\Big(i \gamma^{\mu}(\partial_\mu - i e A_\mu) + m(x) e^{i \phi \gamma^5} \Big)\psi.
}
where $\psi$ is coupled to a string defect defined by the complex field $m(x) \exp(i \phi)$.
$m(x)$ vanishes at the core of the string running along the $z$-axis and $\phi$ winds by $2\pi$ in going around the string.
This mass profile ensures that tere exist chiral fermionic zero modes living on the string which exhibit a gauge anomaly through their coupling to $A_\mu$.

However, integrating out the massive bulk modes provides a Wess-Zumino term \WessYU\ whose gauge variation compensates for the lack of gauge invariance of the chiral zero modes alone.
This can be understood through the following calculation.
Consider a region of spacetime away from the string core where $m(x)$ is non-zero and let us integrate out $\psi$.
The correction to the action by the massive bulk fermions can be inferred by integrating the one-loop correction to the current expectation value,
\eqn\current{\langle J_\nu \rangle = {e \over 8 \pi^2} \epsilon_{\mu \nu \rho \sigma} \partial_\mu \phi F_{\rho \sigma}.
}
Because $\phi$ winds by $2 \pi$ when encircling the string, it formally obeys the equation,
\eqn\eomphi{(\partial_x \partial_y - \partial_y \partial_x) \phi = 2 \pi \delta(x) \delta(y).}
Thus, \current\ implies the bulk fermion addition to the current conservation equation,
\eqn\currentadd{\partial^\mu \langle J_\mu \rangle = {e \over 4 \pi} F_{t z} \delta(x) \delta(y).}
The current is conserved away from the string and the right hand side of \currentadd\ is non-zero along the string so as to cancel the contribution from the chiral zero modes.
This is summarized by a correction to the effective action,
\eqn\wz{S_{WZ} = {e^2 \over 16 \pi^2} \int d^4 x \epsilon_{\mu \nu \rho \sigma} \partial_\mu \phi A_\nu F_{\rho \sigma},}
whose variation under a gauge transformation cancels with the anomalous variation of the chiral zero mode action.
The cancellation between boundary and bulk anomalies is called anomaly inflow as the direction of current flow is towards the boundary.

We stress that the contribution of the massive bulk states is {\it added} to the low energy effective action for the string defect.
We shall not find this prescription to be strictly valid for a domain wall defect.

In order to describe a domain wall, we merely change the soliton profile.
$m(x)$ now vanishes along the domain wall and $\phi$ jumps from 0 to $\pi$ in moving through the wall.
Naively, precisely the same calculation of integrating out the fermions in a region where $m(x)$ is non-vanishing gives the contribution to the effective action \wz.

However, this reasoning is in fact incorrect.  The essential difference is that $\phi$ is constant everywhere except in a small neighborhood of the domain wall where $m(x)$ vanishes.
This is to be contrasted with the string case where $\phi$ wound by $2\pi$ around the string.
Because $\phi$ is constant in the region of space where the calculation obtaining \wz\ is valid, this correction vanishes everywhere that the calculation is well-defined.

We can make the calculation well-defined in the vicinity of the domain wall if we smooth out the field $\phi$. 
Instead of jumping discontinuously at the location of the domain wall, we allow it to smoothly interpolate between zero and $\pi$. 
In effect, this smooth interpolation imparts a non-zero $P$ and $T$ breaking mass that we denoted by $\sigma$ in the main text.
As we have understood through more direct means, this merely breaks $P$ and $T$ on the defect and results in a level 1/2 CS term on the domain wall.

The discontinuous limit is not strictly available within the above scheme. 
This limit is equivalent to taking $\sigma \rightarrow 0$.
Indeed, integrating out the fermions is not well defined in this limit when the mass of the localized fermions vanishes and so the $\sigma \rightarrow 0$ limit need not commute with this integration.
Instead, if we are interested in vanishing bound state mass, we first take $\sigma \rightarrow 0$ and then integrate out the fermions to obtain the 1PI effective action.
This action does {\it not} contain a CS term.

\appendix{C}{The Parity Anomaly and Strong Topological Insulators}

In this appendix, we briefly recall the relationship between the presence of time-reversal protected gapless surface states and the parity anomaly in strong topological insulators.

This relationship is most apparent from the definition of a STI proposed by \refs{\FuKanePump,\LevBur}. 
They consider the fate of a STI on the thickened spatial torus $S^1 \times S^1 \times I$, which has two disconnected toroidal surface boundaries, and two non-contractible curves through which we may insert magnetic flux. 
There are four flux choices for which the system is time-reversal invariant $(\Phi_1, \Phi_2) = (0,0), (\pi,0), (0, \pi),$ and $(\pi, \pi)$.   
If the many-body ground state is Kramers degenerate (i.e., orthogonal to its time-reversed conjugate, which is necessarily a state of the same energy) in an odd number of the 4 flux sectors, the surface spectrum has 2+1 dimensional gapless surface states (in the thermodynamic limit)\LevBur.  
These gapless states cannot be eliminated without breaking the Kramers degeneracy, and therefore breaking time-reversal symmetry; hence the system is a STI.  
This definition is equivalent to the band-structure based definition of \refs{\fukanemele, \moorebalents,\roy} in the non-interacting case, but has the advantage that it does not require an explicit knowledge of the band structure, and thus is equally applicable to interacting systems.   

This criterion ensures that,
if time-reversal symmetry is preserved, there must be a large gauge transformation in which at each surface an odd number of fermions cross from below the Fermi surface to above it.  
Let us begin in a flux sector where the ground state is non-degenerate.  Next, we adiabatically insert flux to arrive in a flux sector where the many-body ground state is Kramers degenerate.  
Applying $T$ (which maps the many-body ground state to its orthogonal Kramers partner), and inserting the same flux again must return the system to its original many-body ground state (since this is equivalent to inserting no flux at all).  If we simply apply the flux insertion twice, without performing a time-reversal transformation in between, 
 we therefore obtain a fermionic configuration that is orthogonal to the original.  As the original many-body ground state was non-degenerate, after this large gauge transformation the system must be in an excited state.  If the only zero-energy state encountered during the flux insertion is in the Kramers doublet, then only one fermionic mode crosses the Fermi surface. (More generally, extra band crossings not protected by Kramers theorem can occur in pairs at momenta $(p, -p)$, and an odd number of fermions can be transferred.)  Hence, there exists a large gauge transformation in which an odd number of fermions cross the Fermi surface.

\appendix{D}{1+1d Effective Action Calculation}

In this section, we repeat the calculation of \S4 in 1+1d. 
This is technically simpler than the higher-dimensional calculation and so is easier to follow.
Our results are consistent with those in \S3 and also allows us to make direct contact with the work of Goldstone and Wilczek \GoldstoneKK.

We begin with the 1+1d action,
\eqn\startaction{S = \int d^2x\ \bar{\psi}\Big(i \gamma^{\mu}(\partial_\mu - i e A_\mu) + m(x) + i \gamma^5 \sigma \Big)\psi,
}
where $\bar{\psi} = \psi^{\dagger} \gamma^0$, $\sigma$ is a constant, and $\lim_{x \rightarrow \pm \infty} = \pm m_0$ for $m_0 > 0$ with a single zero crossing at $x=0$. 
Any such soliton configuration in the mass is stable in 1+1d as opposed to higher dimensions.
If $\sigma \neq 0$, we may safely integrate out $\psi$ without any IR divergences.
This produces an effective action for $A_\mu$ at energies less than $\sigma$.
The leading term is linear in $A_\mu$ and we expect it to take the form,
\eqn\effhypoth{S_{{\rm eff}} = - e \int d^2 x\ A_\mu \tr \Big(\gamma^\mu D(x,x)\Big),
}
where $D(x,x')$ is the propagator for $\psi$ evaluated at $x=x'$ (note that $D(x,x')$ is a $2 \times 2$ matrix and so the trace does not automatically vanish).
Without resorting to an argument \refs{\GoldstoneKK, \CallanSA} presented in Appendix B, it is necessary to construct the real space propagator because there is no translation invariance in the direction perpendicular to the domain wall.

Thus, we must calculate $D(x,x')$.
First, we choose the following representation for the Dirac matrices,
\eqn\diracslow{\gamma^0 = \sigma^1,\quad \gamma^1 = i \sigma^3,\quad \gamma^5 = \gamma^0 \gamma^1 = \sigma^2.
}
The propagator of the Dirac fermion is determined by the eigenvalues and the eigenfunctions of the equation,
\eqn\eigeneqn{\gamma^0 \Big(i \gamma^\mu \partial_\mu + m(x) + i \gamma^5 \sigma \Big) \psi^\lambda = \lambda \psi^\lambda,
}
which can be written as
\eqn\eigeneqnmat{\pmatrix{i \partial_t - \sigma && \partial_x + m(x) \cr - \partial_x + m(x) && i \partial_t + \sigma} \pmatrix{\psi_1^{\lambda} \cr \psi_2^{\lambda}} = \lambda  \pmatrix{\psi_1^{\lambda} \cr \psi_2^{\lambda}}.
}
The propagator is formally given by the expression,
\eqn\propagator{D(x,x') = i \sum_\lambda {\psi^{\lambda}(x) \bar{\psi}^{\lambda}(x') \over \lambda},
}
where the $\psi^\lambda$ are normalized eigenfunctions.

Time-translation invariance allows us to Fourier expand $\psi_i^{\lambda}(t,x) = \int_\omega \psi_i^{\lambda}(w,t) \exp(- i w t)$.
Thus, we must solve the following eigenvalue problem,
\eqn\eigeneqnmat{\pmatrix{\omega - \sigma && \partial_x + m(x) \cr - \partial_x + m(x) && \omega + \sigma} \pmatrix{\psi_1^{\lambda}(\omega, x) \cr \psi_2^{\lambda}(\omega, x)} = \lambda  \pmatrix{\psi_1^{\lambda}(\omega, x) \cr \psi_2^{\lambda}(\omega, x)}.
}
Spatial translation invariance is restored as $x \rightarrow \pm \infty$. 
We can use this fact to solve for the eigenvalues (of any states that are not localized to the domain wall) by asymptotically Fourier decomposing, $\lim_{x \rightarrow \pm \infty} \psi_i^{\lambda}(\omega, x) = \int_k \psi_i^{\lambda}(\omega, k) \exp(- i k x)$. 
\eigeneqnmat\ becomes
\eqn\eigeneqnmatinfty{\pmatrix{\omega - \sigma && -i k \pm m_0 \cr i k \pm m_0 && \omega + \sigma} \pmatrix{\psi_1^{\lambda}(\omega, k) \cr \psi_2^{\lambda}(\omega, k)} = \lambda  \pmatrix{\psi_1^{\lambda}(\omega, k) \cr \psi_2^{\lambda}(\omega, k)}.
}
Thus, there are two sets of eigenfunctions, $\psi^{\pm}(\omega, x)$, with eigenvalues, $\lambda_{\pm} = \omega \pm \sqrt{k^2 + m_0^2 + \sigma^2} = \omega \pm \omega_k$.
In addition to these scattering states, there are modes bound to the domain wall as we shall describe.

To find the eigenfunctions of the scattering states, we must solve the equation,
\eqn\eigeneqnmatfinal{\pmatrix{\omega - \sigma && \partial_x + m(x) \cr - \partial_x + m(x) && \omega + \sigma} \pmatrix{\psi_1^{\pm}(\omega, x) \cr \psi_2^{\pm}(\omega, x)} = \lambda_\pm  \pmatrix{\psi_1^{\pm}(\omega, x) \cr \psi_2^{\pm}(\omega, x)}.
}
It is possible to do this by finding the eigenfunctions for the constant mass Dirac equation and then replacing the constant mass by the spatially varying one. However, the following observation is instructive.
The coupled first-order equation \eigeneqnmatfinal\ may be written as two decoupled second order equations,
\eqn\secdirac{\eqalign{\Big(\partial_x^2 - \partial_x m(x) - m(x)^2 + m_0^2 + k^2 \Big)\psi_1^{\pm} = & 0 \cr
\Big(\partial_x^2 + \partial_x m(x) - m(x)^2 + m_0^2 + k^2 \Big)\psi_2^{\pm} = & 0.
}}
To proceed further, we must choose a particular form of the interpolating mass.
We take $m(x) = m_0 \tanh(x/\ell)$.
Substituting this into \secdirac\ we find
\eqn\secdiractwo{\eqalign{\Big(\partial_x^2 + k^2 + (1 - 1/m_0 \ell){m_0^2 \over \cosh(x/\ell)} \Big)\psi_1^{\pm} = & 0 \cr
\Big(\partial_x^2 + k^2 + (1 + 1/m_0\ell){m_0^2 \over \cosh^2(x/\ell)} \Big)\psi_2^{\pm} = & 0.
}}
These are generalized P\"oschl-Teller equations.

There is a well known relation between the P\"oschl-Teller equations and supersymmetric quantum mechanics as we now describe \stonegoldbart.
For simplicity, let us work at $\ell = 1/m_0$.
Define the (supersymmetric charge) operator,
\eqn\charge{Q = \partial_3  - m_0 \tanh(m_0 x_3), \quad Q^{\dagger} = - \partial_3  - m_0 \tanh(m_0 x_3).}
We may suggestively rewrite \secdiractwo\ so that the equation satisfied by each eigenspinor component takes one of two forms: 
\eqn\pteqn{Q^\dagger Q (\psi^{\lambda_{\pm}}_k)_1 = (\omega_k^2 - \sigma^2) (\psi^{\lambda_{\pm}}_k)_1, \quad Q Q^{\dagger} (\psi^{\lambda_{\pm}}_k)_2 = (\omega_k^2 - \sigma^2) (\psi^{\lambda_{\pm}}_k)_2,
}
where the subscript, $1,2$, refers to the spinor component.
Recall that the eigenvalue in the above equation $(\omega_k^2 - \sigma^2) = k^2 + m_0^2$.

Scattering states are defined to be those eigenfunctions for which $(\omega_k^2 - \sigma^2) \neq 0$.
There is a pairing between scattering state spinor components;
given a solution $\psi_1$ to the first equation in \pteqn\ with non-zero eigenvalue, a solution to the second equation with identical eigenvalue is given by $Q \psi_1$.
A bound state is annihilated by either $Q$ or $Q^\dagger$ and there is no corresponding pairing of spinor components. 
Consequently, bound states can only become scattering states in pairs.
The number of bound states is given by the index of $Q$; this number is the difference in dimensions of the kernel (or null space) of the operators $Q$ and $Q^\dagger$.
For our problem, only states annihilated by $Q^\dagger$ are normalizable and so we need only consider this operator when finding the bound state wave function.
The number of bound states or zero modes is preserved under small changes of $\ell$.
For $0 < m_0 \ell \leq 1$, there exists a single zero mode bound state.

Following this brief digression, let us now directly solve \secdiractwo\ at the point $1/\ell = m_0$ for which the two equations simplify to
\eqn\secdiracthree{\eqalign{\Big(\partial_x^2 + k^2 \Big)\psi_1^{\pm} = & 0 \cr
\Big(\partial_x^2 + k^2 + {2 m_0^2 \over \cosh^2(m_0 x)} \Big)\psi_2^{\pm} = & 0.
}}
The equation for $\psi_1^{\lambda_{\pm}}$ is easily solved by ${\rm const.} \exp(- i k x)$.
It may be substituted back into the first order differential equations \eigeneqnmatfinal\ in order to determine $\psi_2^{\lambda_{\pm}}$.
Doing so, we find the un-normalized eigenspinors for the scattering states,
\eqn\eigenfunctions{\psi^{\lambda_{\pm}}_{\omega, k}(t, x) = \pmatrix{\psi_1^{\pm}(t, x) \cr \psi_2^{\pm}(t, x)} = \pmatrix{\pm \omega_k - \sigma \cr i k + m_0 \tanh(m_0 x)} e^{- i w t - i k x}.
}
Note that $k$ only has the strict interpretation of momentum asymptotically far away from the domain wall when $m(x)$ becomes a constant; otherwise, it may be viewed as a reparameterization of the eigenvalues.  
It is interesting that $\sigma$ never explicitly enters the second order differential equations \secdiractwo.

It remains to find the bound state.
The two components of the Dirac spinor are no longer paired. 
Instead, $\psi_2$ is the only non-zero component of the spinor and we need only solve the first order differential equation \eigeneqnmatfinal\ with $m(x) = m_0 \tanh(m_0 x)$.
(There is no corresponding normalizable solution where $\psi_2 = 0$ and $\psi_1$ is non-zero.)
The un-normalized solution is 
\eqn\zeromd{\psi^{\lambda_0}_{\omega}(t,x) = \pmatrix{0 \cr \sech(m_0 x)} e^{- i \omega t},}
with eigenvalue $\lambda_0 = \omega + \sigma$.
It is a bit of a misnomer to refer to $\psi^{\lambda_0}$ as a zero mode as its energy is bounded from below by $\sigma$. 
It is better to call it a bound state.
Of course, this bound state is free to mix with the bulk continuum when its energy is comparable to the bulk gap.

Equipped with the eigenspinors \eigenfunctions\ and \zeromd, we must now normalize them with respect to the inner product \innerprod\ where we replace the number of integration dimensions by two and $n= 1,2$ depending upon whether the state is localized or extended.
Thus, the orthonormal eigenspinors are (same notation as un-normalized ones above)
\eqn\spinorsfinal{\eqalign{\psi^0_\omega(t, x) = & \sqrt{{m_0 \over 4 \pi}} \pmatrix{0 \cr \sech(m_0 x)} e^{- i \omega t}, \cr
\psi^{\pm}_{\omega, k}(t, x) = & \sqrt{{1 \over 8 \pi^2 \omega_k (\omega_k \mp \sigma)}}\pmatrix{\pm \omega_k - \sigma \cr i k + m_0 \tanh(m_0 x)} e^{- i w t - i k x}.
}}

Let us now construct the propagator using \propagator.
This is a $2 \times 2$ matrix. 
The sum over the eigenvalues becomes integrals over $\omega, k$.
The zero mode only contributes a non-zero $21$-entry (given our choice of $\gamma$-matrices),
\eqn\zeropropagator{D^{{\rm bound}}(x,x') = {m_0 \over 4} (1 + i \gamma^1) \sech(m_0 x) \sech(m_0 x') \int {d \omega \over 2 \pi} {\gamma^0 \omega + i \gamma^5 \sigma \over \omega^2 - \sigma^2} e^{i \omega (t' - t)}.
}
Note that this is simply the propagator of a massive 0+1d particle. 

The contribution to the propagator from the scattering states is more complicated, but straightforwardly found as before,
\eqn\scatpropconcise{D^{{\rm scat}}(x, x') = \int {d \omega dk \over (2 \pi)^2} {\gamma^\mu k_\mu + M(x,x') + i \gamma^5 \sigma \over \omega^2 - k^2 - m_0^2 - \sigma^2}e^{i \omega(t' - t) + i k (x' - x)},
}
\eqn\massmatrixcompact{M(x,x') = - {1 \over 2} (1 - i \gamma^1) m(x') - {1 \over 2} (1 + i \gamma^1) m(x) + {1 \over 2} (1 + i \gamma^1) {\omega \gamma^0 + i \sigma \gamma^5 \over k^2 + m_0^2} \mu(x,x')
}
and
\eqn\another{\mu(x, x') = m(x) m(x') - i k (m(x) - m(x')) - m_0^2.
}
The similarity between the two-dimensional and four-dimensional cases is evident. 

Return to the expression for the leading contribution to the effective action for $A_\mu$,
\eqn\effhypoth{S_{{\rm eff}} = - e \int d^2 x\  A_\mu \tr \Big(\gamma^\mu D(x,x)\Big).
}
In 1+1d, we only need the propagator evaluated at the same starting and ending point. 
Therefore, let us simply let $t'=t$ and $x'=x$ in our expressions for the bound and scattering state propagators.
Using our expression for the propagator at coincident points, we find
\eqn\efftotal{S_{{\rm eff}} = \Big({\eta \sgn(\sigma) \over 2}
- {1 \over \pi}  \tan^{-1}({\sigma \over \eta m_0}) \Big) \int dt\ e A_0,
}
where $\eta = \pm 1$ allows for a general domain wall orientation, $m(x) = \eta m_0 \tanh(m_0 x)$.
We also used the relation,
\eqn\approx{\lim_{m_0 \rightarrow \infty} {m_0 \sech^2(m_0 x) \over 2} = \delta(x).}
\efftotal\ is simply the 0+1d CS term for the $A_0$ field.
In the limit of interest, $\sigma/m_0 \rightarrow 0$, the level is half-integral.
Notice that the total level depends upon the asymptotic ratio $\sigma/m_0$ precisely in the way predicted by Goldstone and Wilczek \GoldstoneKK.
While our calculation was performed for a particular soliton profile, we expect the level to be $1/2\pi$ times the change in phase of the bulk fermion mass.

\appendix{E}{Bulk and Boundary Corrections}

Here we will give the details of the calculation of the $1$-loop correction to the gauge field effective action.  
This correction takes the form,
\eqn\actionefffour{\eqalign{\delta S = & \int d^4 x d^4 x' A_\mu(x) \Pi^{\mu \nu}(x, x') A_\nu(x') \cr
= & {e^2 \over 2} \int d^4 x d^4 x' A_\mu(x) A_\nu(x') \tr\Big(\gamma^\mu D(x,x') \gamma^\nu D(x',x)\Big).
}}
In \S4, we studied in detail the contribution of $D^{{\rm loc}}$ to \actionefffour.
Because of the softer UV behavior of this propagator, we found this term to be finite in the UV cutoff.
Thus the level $1/2$ CS term arising from the fermion propagator is, unlike in a $2+1$d system, unmodified by any regularization procedure.

It is important to verify that any divergences in the $\Lambda \rightarrow \infty$ limit arising from other contributions to \actionefffour\ do not force us to add regulator fields that will change the level of the Chern-Simons coefficient modulo $1$.  
To verify this, we write the 3+1d propagator as the sum of two terms,
\eqn\propfull{D(k, x_3, x_3') = D^{{\rm loc}}(k, x_3, x_3') + D^{{\rm free}}(k, x_3, x_3'),
}
where
\eqn\proploc{D^{{\rm loc}}(k,x_3, x_3') = {\mu(x_3, x_3')\mu_{ i k_3 \ {\rm sign}(x_3'-x_3)} \over 4} (1 + i \gamma^3)  {\gamma^a k_a + i \sigma \gamma^5 \over (-k_a^2 + \sigma^2)\sqrt{-k_a^2 + m_0^2 + \sigma^2}} e^{- k_3 |x_3' - x_3|},
}
and
\eqn\Dfree{
D^{\rm{free}}(k, x_3, x_3')  = { \gamma^a k_a -  \gamma^3 \kappa - {1 \over 2} \left[  m(x_3) + m(x_3')  + i \gamma^3(m(x_3)-  m(x_3') ) \right ]  + i \sigma \gamma_5 \over{ 2 \sqrt{-k_a^2 + m_0^2 + \sigma^2}} } e^{- k_3 |x_3' - x_3|}
}
for $k_3 = \sqrt{- k_a^2 + m_0^2 + \sigma^2}, \kappa=i k_3 {\rm sign} (x_3' - x_3 )$.   Here we have  integrated over all $k_3$ in \scatpropthree, such that the contribution from the poles at $k_3 = \pm i m_0$ exactly cancels the propagator of the bound states, as explained in Sect. 4.  
This gives $\Pi_{\mu \nu} =\Pi_{\mu \nu}^{\rm{f, f}}+ \Pi_{\mu \nu}^{\rm{f, l}}  + \Pi_{\mu \nu}^{\rm{l, l}}$, with
\eqn\OneLoops{ 
\Pi_{\mu \nu}^{\rm{f,f}} = \rm{Tr} \left( \gamma_\mu D^{\rm{free}} \gamma_\nu D^{\rm{free}} \right ) 
\ \ \ \ 
 \Pi_{\mu \nu}^{\rm{l, l}} = \rm{Tr} \left( \gamma_\mu D^{\rm{loc}} \gamma_\nu D^{\rm{loc}} \right ) 
 }
 \eqn\OLTwo{\Pi_{\mu \nu}^{\rm{f, l}} = \rm{Tr} \left( \gamma_\mu D^{\rm{free}} \gamma_\nu D^{\rm{loc}} +    \gamma_\mu D^{\rm{loc}} \gamma_\nu D^{\rm{free}}   \right ).
 }
 In the main text we discussed only $\Pi_{\mu \nu}^{\rm{l, l}}$; here we will evaluate the remaining contributions.  

To understand the structure of the possible divergences, it is helpful to Fourier transform in $x_3$.
Recall that the assumed translation invariance in the directions tangent to the domain wall allow us to immediately work in a mixed Fourier basis where $k_a$ for $a = 0, 1, 2$ is the momentum tangent to the domain wall.
Focusing on the third spatial direction, we introduce the momenta $s, t$:
\eqn\Ft{\eqalign{
 \int d x_3 \ d x_3'  \Pi_{\mu \nu}(x_3, x_3', \vec{q}_a) A_\mu(x_3, \vec{q}_a) A_\nu(x'_3, -\vec{q}_a ) \cr   =\int d x_3 \ d x_3'  \int d s \  d t \ e^{ i s x_3 } e^{ i t x'_3} A_\mu(s, \vec{k}_a) A_\nu (t, \vec{k}_a)  \Pi_{\mu \nu}(x_3, x_3', \vec{k}_a) 
}}
Intuitively, this form is convenient because fixing $x_3, x_3'$ at the location of the domain wall requires us to include modes of arbitrarily high energy, obscuring the structure of the divergences.  
Thus, it is preferable to first integrate over $x_3, x_3'$ and then examine the structure of the divergences for small $s, t, \vec{q}_a$.  

As a sanity check, it is instructive to see what this prescription gives for the case of a translationally-invariant mass.  
In this case, the only $x_3$ dependence of $\Pi_{\mu \nu}$ is  
$$\Pi_{\mu, \nu} (x_3, x_3', \vec{q}_a) = \int d^3 \vec{k}_a { e^{ -( k_3+ k_3') |x_3' - x_3 |}  \over k_3 k'_3} \left(( k \cdot k' - m^2)  \eta_{\mu \nu} -  k_\mu k'_\nu - k'_\mu k_\nu \right ),$$ 
where $k = (k_0, k_1, k_2, k_3), \ k' = (k_0-q_0, k_1 - q_1, k_2-q_2, k_3)$   
and
\eqn\kthrees{
k_3 = \sqrt{ -k_0^2+ k_1^2+k_2^2+m_0^2+\sigma^2}, \quad
k_3' = \sqrt{ -(k_0 - q_0)^2+ (k_1 - q_1)^2+(k_2 - q_2)^2+m_0^2+\sigma^2}.
}
In this case, performing the integral over the center-of-mass co-ordinate $x_3 + x_3'$ enforces the condition $s = - t$.  
The remaining integral over $x_3' - x_3$ gives:
\eqn\zints{
\int d (x_3'- x_3) \Pi_{\mu \nu}(x_3, x_3', \vec{p}_a)  =  \int d^3 \vec{k}_a \ 2{k_3 + k_3'  \over (k_3 + k_3')^2 + s^2} {1  \over k_3 k'_3} \left(( k \cdot k' - m^2)  \eta_{\mu \nu} -  k_\mu k'_\nu - k'_\mu k_\nu \right). 
  }
As one might expect, this is exactly the result we would obtain by first Fourier transforming the propagator in all four spacetime momenta, and then integrating over $k_3$, with $q_3 = s$.  

When the mass breaks translation invariance in $x_3$, $s$ and $t $ are not conserved, and integrating over $x_3, x_3'$ will not force $s = -t$.  
However, it remains true that $\Pi_{\mu \nu} (x_3, x_3', \vec{q_a})$ is an admixture of modes at many different energies, and that we must integrate over $x_3$ and $x_3'$ in order to obtain the correct structure of divergences.  

Defining  $z = x_3' - x_3$, $Z = x_3 + x_3'$, the various contributions to the propagator are:
\eqn\Pimumu{ \eqalign{
\Pi_{\mu \nu}^{{\rm f,f}}( q_a,x_3, x_3') =& \int d^3 k_a {e^{ - (k_3 + k_3') |z| } \over  k_3 k_3'} \left[  (  k^\rho  (k_\rho+ q_\rho) - m_0^2 - \sigma^2 ) \eta_{\mu \nu} - k_\mu (k_\nu + q_\nu)  - k_\nu (k_\mu + q_\mu)  \right . \cr
& \left. +  \ \eta_{\mu \nu} T_0  - T_1  \left( i \gamma^z  + i {1 \over ( k_3 + k_3') \ {\rm sign } (z) }  \delta_{\nu a} q_a  \delta_{\nu 3} \right ) - \delta_{\mu 3} \delta_{\nu 3} {2 T_1^2 \over ( k_3 + k_3')^2 } \right ] \cr
\Pi_{\mu \nu}^{{\rm f,l}}( q_a,x_3, x_3') =& \int d^3 k_a {e^{ - (k_3 + k_3') |z| } \over  k_3 k_3'} T_{{\rm cross} }\left[  \delta_{\mu a} \delta_{\nu b} \left \{ (  k^c  (k_c + q_c)   - \sigma^2 )  \eta_{a, b} - k_b (k_a + q_a)   \right . \right . \cr
&  \left . \left . - k_a (k_b + q_b)  + i \sigma \epsilon_{ a,b,c} q_c \right \} - \delta_{\mu 3} \delta_{\nu 3} \left(  k^c  (k_c + q_c) - \sigma^2 \right ) + \alpha
\right ]  \cr
\Pi_{\mu \nu}^{{\rm l,l}}( q_a,x_3, x_3') =&\int d^3 k_a {e^{ - (k_3 + k_3') |z| } \over  k_3 k_3'} T_{{\rm loc} }\left[  \delta_{\mu a} \delta_{\nu b} \left \{ (  k^c  (k_c + q_c)   - \sigma^2 )  \eta_{a, b} - k_b (k_a + q_a) 
 \right . \right . \cr &  \left . \left .  
 - k_a (k_b + q_b) + i \sigma \epsilon_{ a,b,c} q_c \right \}  \right ] 
}
}
where we have defined $k_3' = \sqrt{ - (k_0 + q_0)^2 + (k_1 +q_1)^2+(k_2+q_2)^2+m_0^2+\sigma^2}$.  Here $\alpha$ is a matrix of linear order in $q_a$ that is non-zero only in the fourth row and column, and 
\eqn\TheTs{ \eqalign{
T_0 =&  -{m_0^2 \over 2}  \left( \tanh^2 m_0 x_3 +\tanh^2 m_0 x_3' \right ) + m_0^2  = 
 2 m_0^2 { 1 + \cosh m_0 Z  \cosh m_0 z \over ( \cosh m_0 Z + \cosh m_0 z)^2 }  \cr
T_1 = & {m_0 (k_3 + k_3') {\rm sign}( z) \over 2 } \left( \tanh m_0 x_3' - \tanh m_0 x_3 \right )  =  
 { m_0 (k_3 + k_3')   \sinh m_0 | z   |\over \cosh m_0 Z + \cosh m_0 z  }\cr
T_2 = & m_0^2 ( 1 -  \tanh m_0 x_3  \tanh m_0 x_3') 
= { 2 m_0^2 \cosh z\over \cosh m_0 Z + \cosh m_0 z  } \cr
T_{{ \rm cross}} =&  
    {1 \over  -k_a^2+ \sigma^2 } (T_1 +T_2 ) + O(q_a ) \cr
T_{{ \rm loc}} 
=&  {1 \over 2 ( - k_a^2+ \sigma^2 )^2 } (T_1 +T_2 )^2 + O(q_a )
}
}

For $k^2$ large, $\Pi_{\mu \nu}^{\rm{f, f}}$ and $\Pi_{\mu \nu}^{\rm{f, l}}$ scale like $|k|^{-1}$, while $\Pi_{\mu \nu}^{\rm{l, l}} $ scales like $|k|^{-2}$ -- hence naively, after integrating over the remaining three loop momenta, all three might be divergent.  However, care must be taken with this naive power-counting, as is apparent from the form of $\Pi_{\mu \nu}^{\rm{QED4} }$ prior to integrating over $z,Z$: power-counting suggests that the leading-order divergence should be cubic, while it is in fact quadratic in $\Lambda$.  
Likewise, we will find that the leading-order divergence from the terms that arise due to the spatial variation of $m(x_3)$ is logarithmic.  

To exhibit these divergences explicitly, we next Fourier transform in $x_3$, as in Eq. \Ft ,  and perform the integrals over $z$ and $Z$.  For example, we wish to integrate 
\eqn\Zints{
{1 \over 2} \int d Z \ dz e^{ -2  k_3 |z| } e^{ i (s + t)/2 Z} e^{i (s-t)/2 z} T_0.
}
To do this for general $s$ and $t$, we observe that the zeros in the denominator of $T_0$ occur at $Z = \pm z + i (2 n+1 ) \pi$.  Summing over $n$, we obtain
\eqn\ZintsT{\eqalign{
{1 \over 2} \int d Z \ e^{ -2  k_3 |z| } e^{ i (s + t)/2 Z} e^{i (s-t)/2 z} T_0 = { 1 \over 2 \sinh {\pi (s+t) \over 2 m_0} }  e^{ -2  k_3 |z| }  e^{i (s-t)/2 z} \cr \left [ \rm{Res}(T_0; x = y + i \pi) + \rm{Res}(T_0; x = y - i \pi) \right ].
} }
We then integrate these expressions with respect to $z$.  
After performing both integrations,  the mass dimension of the result has been reduced by $2$ (rather than by $1$, as it is when integrating over $Z$ leads to $\delta(s - t)$).  This indicates that the corresponding contribution to $\Pi_{\mu \nu}$ has the mass dimension of the correction for a $2+1$D theory.  

We obtain:
\eqn\FPa{
m_0^2 \int d z \ d Z e^{ -2  k_3 |z| } e^{ i (s + t)/2 Z} e^{i (s-t)/2 z} T_0 = 2 \pi {\rm sign} (m_0) { k_3 (s + t) ( k_3^2 +{1 \over 8} ( s^2 + t^2)) \over ( e^{\pi (s + t)/
   m_0} -1) ( k_3^2 +{1\over 4}  s^2) ( k_3^2 +{1\over 4}  t^2)},
   }
The exponential $e^{\pi (s + t)/
   m_0}$  ensures that we may expend the result for small $s,t \over k_3$: not only do these parametrize the photon momentum in the $x_3$ direction, which we take to be small relative to $m_0$, but equally the result is exponentially suppressed in ${s + t \over m_0}$, while after Wick rotating, $k_3 \geq \sqrt{m_0^2+ \sigma^2}$.  
   
This expansion can be safely performed for all of the integrals.  To zeroeth order in $s,t$, we obtain:   
\eqn\junk{ \eqalign { 
{1 \over 2} \int d z d Z e^{ - 2 k_3 |z| } T_0 = & {2 |m_0| \over k_3 } \ \ \ \ \ \ \ \ \ \ \ \ \ \ \ \ \
{1 \over 2}\int d z d Z e^{ - 2 k_3 |z| } T_1= {  |m_0| \over k_3 } \cr
{1 \over 2}\int d z d Z e^{ - 2 k_3 |z| } T_2=&\left ( -{  m_0^2 \over k_3^2} + 2  |m_0| \partial_{k_3} \Gamma( {k_3 \over |m_0|} ) \right )  \cr
{1 \over 2}\int d z d Z e^{ - 2 k_3 |z| } T_1^2 =& -2  \left [m_0^2 + 2 k_3 | m_0| - 2 k_3^2 |m_0| \partial_{k_3} \Gamma ({k_3 \over |m_0|} )  \right ] \cr
{1 \over 2}\int d z d Z e^{ - 2 k_3 |z| } T_2^2 = &2  (2 k_3^2 + m_0^2) \left [ 
    {-2 k_3^5 |m_0|+ 7 k_3^4 m_0^2 - 8 k_3^3 |m_0|^3 + k_3^2 m_0^4 + 4 k_3 |m_0|^5 - 
      4 m_0^6 \over k_3^2 (k_3 - 2| m_0|)^2 (k_3 -| m_0|)^2}  \right . \cr
 &  \left.    + 
   2 |m_0| \partial_{k_3}  \Gamma( -2 + {k_3 \over |m_0|}) )\right ] \cr 
{1 \over 2}\int d z d Z e^{ - 2 k_3 |z| }  T_1 T_2 = &  2 \left [ {|m_0| (4 k_3^4 - 2 k_3^3 |m_0| + k_3^2 m_0^2 + m_0^4) \over k_3 (k_3 - |m_0| )^2 } - 
   4 k_3^2 |m_0| \partial_{k_3} \Gamma(-1 +{k_3 \over |m_0|} )) \right ]
} }
where $\Gamma$ is the digamma function.  

Taylor expanding these expressions for large $k_a$, one can see that $\Pi_{\mu \nu}^{{\rm l,l}}( 0,s=0, t=0)$ is non-divergent, as claimed in Sect. 4.  Further, the divergent terms associated with the domain wall in $\Pi_{\mu \nu}^{{\rm f,f}}( 0,s=0, t=0)$ and $\Pi_{\mu \nu}^{{\rm f,l}}( 0,s=0, t=0)$ cancel.

We can also evaluate the extra finite contributions to the gauge non-invariant terms due to the contributions we neglected in Sect. 4.  Inserting the expressions in \junk\ into the expression for the Fourier-transformed propagator, and Wick rotating to send $k_a^2 \rightarrow - k_a^2$, we obtain:
\eqn\PiTot {
\Pi_{\mu \nu} (0,0,0) =\Pi_{\mu \nu} (0,0,0)^{QED4}-  \int d^3 k_c { 2 m_0 (m_0^2 k_c^2 + 3 \sigma^2  \kappa^2 ) \over 3 \kappa^3 (k_c^2 + \sigma^2)^2 }\delta_{\mu a} \delta_{\nu b} \  \eta_{a b}  = - {2 |m_0| \over 3}\delta_{\mu a} \delta_{\nu b} \  \eta_{a b} 
}
where $\Pi_{\mu \nu} (0,0,0)^{QED4}$ is the gauge non-invariant contribution from the fermion loop in QED4.  

Thus the gauge non-invariant terms that arise due to the domain wall do indeed have the form given in Eq. \locsecond , and regularization can be carried out with an even number of Pauli-Villars fields with spatially varying masses.

We can also use this method to calculate the coefficient of the Chern-Simons term in Eq. \Pimumu .  As described in Sect. 4, $\Pi_{\mu \nu}^{{\rm l,l}}$ contributes a half-integral Chern-Simons term, whose sign is determined by the sign of the domain-wall mass $\sigma$.  However, there is also a {\it non-quantized} contribution from $\Pi_{\mu \nu}^{{\rm f,l}}$.  
Thus the total effective Chern-Simons term is 
\eqn\csterms{
\Pi_{a b}^{{\rm CS}} = { 1 \over 8 \pi^2}  \left( \pi {\rm sign } ( \sigma )- 2 \sin^{-1} {\sigma \over \sqrt{ m_0^2 + \sigma^2 }  } \right )  \left(  i \epsilon_{a b c} p^c \right)
}
or a Chern-Simons coefficient of 
\eqn\keff{
k = {1 \over 2}  {\rm sign } ( \sigma ) -{  1  \over  \pi}   \tan^{-1} {\sigma \over |m_0| }
 }
where we have used $\sin^{-1} {\sigma \over \sqrt{ m_0^2 + \sigma^2 } } =  \tan^{-1} {\sigma \over |m_0| }$.  
This is reminiscent of the corrections to the induced charge of solitons in the simpler 1+1d model \efftotal: in 3+1d if we break time-reversal (for example by applying a magnetic field to the system) the domain wall carries an induced charge, whose value is exactly quantized only in the limit $\sigma/m_0 \rightarrow 0$.

\listrefs

\end